\documentclass[useAMS,usenatbib]{mn2e}
\usepackage{graphics}
\usepackage{epsf}
\usepackage{subfigure}
\usepackage{amsmath}
\usepackage{amssymb}

\def\ni{\noindent}

\begin{document}

\title[Cepheid PC \& AC relations - IV]{Period-colour and amplitude-colour relations in classical Cepheid variables IV: The multi-phase relations}
\author[Ngeow \& Kanbur]{Chow-Choong Ngeow$^{1}$\thanks{E-mail: cngeow@astro.uiuc.edu} and Shashi M. Kanbur$^{2}$
\\
$^{1}$Department of Astronomy, University of Illinois, Urbana-Champaign, IL 61801, USA
\\
$^{2}$Department of Physics, State University of New York at Oswego, Oswego, NY 13126, USA
}

\date{Accepted 2005 month day. Received 2005 month day; in original form 2005 June 15}

\maketitle

\begin{abstract}

The superb phase resolution and quality of the OGLE data on LMC and SMC Cepheids, together with existing data on Galactic Cepheids, are combined to study the period-colour (PC) and amplitude-colour (AC) relations as a function of pulsation phase. Our results confirm earlier work that the LMC PC relation (at mean light) is more consistent with two lines of differing slopes, separated at a period of 10 days. However, our multi-phase PC relations reveal much new structure which can potentially increase our understanding of Cepheid variables. These multi-phase PC relations provide insight into why the Galactic PC relation is linear but the LMC PC relation is non-linear. This is because the LMC PC relation is shallower for short ($\log P < 1$) and steeper for long ($\log P > 1$) period Cepheids than the corresponding Galactic PC relation. Both of the short and long period Cepheids in all three galaxies exhibit the steepest and shallowest slopes at phases around 0.75-0.85, respectively. A consequence is that the PC relation at phase $\sim0.8$ is highly non-linear. Further, the Galactic and LMC Cepheids with $\log P > 1$ display a flat slope in the PC plane at phases close to the maximum light. When the LMC period-luminosity (PL) relation is studied as a function of phase, we confirm that it changes with the PC relation. The LMC PL relation in $V$- and $I$-band near the phase of 0.8 provides compelling evidence that this relation is also consistent with two lines of differing slopes joined at a period close to 10 days. 

\end{abstract}


\section{Introduction}

    In \citet[][hereafter Paper I]{kan04}, we investigated the behavior of period-colour (PC) and also amplitude-colour (AC) relations at the $V$-band maximum, mean and minimum light, as the AC relations are closely related to the PC relations via the following equation, derived in \citet[][hereafter SKM]{sim93} and in Paper I:

    \begin{eqnarray}
    \log T_{max} - \log T_{min} = \frac{1}{10} (V_{min} - V_{max}).
    \end{eqnarray}

    \ni This equation implies that if the PC relation at the maximum (or minimum) light is flat (i.e. the colour is independent of period), then there will be an AC relation at the minimum (or maximum) light \citep[see SKM, Paper I and][hereafter Paper II]{kan04a}. Thus an examination of AC relations can be used to confirm changes to the PC relation. The major assumption used in equation (1) is that temperature, not radius, fluctuations are the dominant cause of light variations, at least at the optical wavelengths \citep{cox80}. In this paper, in addition to the maximum, mean and minimum light, we extend the study of PC and AC relations at all other pulsation phases. Furthermore, the extension to multi-phase relations will also expand our understanding of the pulsational behavior of classical Cepheids. The results of our study justify our methods and reveal new properties of Cepheids in the Galaxy and Magellanic Clouds that can be used to constrain the theoretical Cepheid pulsation/evolution models in future studies.

    Based on the extensive data and large number of Cepheids recently discovered in the Large Magellanic Cloud (LMC), there is strong evidence that the mean light PC relation for the LMC Cepheids is non-linear: there are two PC relations for the short ($\log P<1$) and long ($\log P > 1$) period Cepheids \citep{tam02,tam02a,kan04,san04,kan06,nge05}\footnote{In the rest of this paper, this is what we mean by non-linearity, and the long and short period Cepheids refer to those with $\log P>1.0$ and $\log P<1.0$, respectively.}, respectively. In contrast, the Galactic or the Small Magellanic Cloud (SMC) Cepheids do not show any non-linearity of the mean light PC relation \citep{tam03,kan04}. Therefore, our motivation is to construct and study the PC relations for Galactic, LMC and SMC Cepheids at all phases between zero and one, because the PC relation at mean light is the average of the PC relation at all phases (see Paper I). Thus it is important to understand the properties of the PC relation at other phases. A direct consequence of the non-linear LMC PC relation at mean light is that the LMC period-luminosity (PL) relation at mean light is also non-linear\footnote{See, for example, \citet{mad91} for the basic physics of the Cepheid PL and PC relations.} \citep{tam02,kan04,san04,nge05}. Although the LMC PL relation is extensively used in distance scale studies \citep[see, e.g.,][]{fre01,sah01}, \citet{nge05w} have discussed the implications of the non-linear mean light LMC PL and PC relations in distance scale applications. 

    In Section 2 of this paper, we present the data and the numerical methods used in this study. The multi-phase PC relations are studied in Section 3, and the detailed comparison between the Galactic and LMC PC relations is given in Section 3.1. The multi-phase AC relations are briefly presented in Section 4. Since it is also of great interest to investigate the connections between the PL and PC relations for LMC Cepheids, the multi-phase LMC PL relations are presented in Section 5. Finally, the main conclusion of this paper is summarized in Section 6.


\section{Data and Method}

     The data used in this study are the same as in Paper I for the Galactic and SMC Cepheids, and in \citet[][hereafter Paper III]{kan06} for the LMC Cepheids. These data include the periods, $V$- and $I$-band photometric data, and the $E(B-V)$ values that have been published elsewhere (see the reference in Paper I \& III). To construct the multi-phase PC \& AC relations, the light curves of the Cepheids should be phased to a common starting epoch. Briefly, the $V$- and $I$-band photometric data are fitted with the $M^{th}$-order Fourier expansion of the following form \citep{sch71,sim81,nge03}:

     \begin{eqnarray}
       m(t) & = & A_0 + \sum^{M}_{k=1} [A_k\cos(2\pi k\Phi(t) + \phi_k)], k=1,...M 
     \end{eqnarray} 

     \ni where $\Phi(t)=(t-t_0)/P - \mathrm{int}[ (t-t_0)/P ]$, with $t_0$ being a common starting epoch for all Cepheids in both bands (say, $t_0=2440000.0$). The pulsation period $P$ (in days) are taken from the literature. The phases $\Phi(t)$ are in between zero and one corresponding to one full pulsation cycle. The Fourier coefficients, $A_k$ and $\phi_k$, as well as the mean values ($A_0$) for the Cepheids in our samples are given in the Appendix A. 

     To obtain the colours at various phases, the phased light curves from equation (2) should be shifted such that phase zero corresponds to ($V$-band) maximum light for all Cepheids in the sample. This was done using the following steps. The phase that corresponds to maximum light\footnote{We assume the $V$-band light curve traces the luminosity curve, hence the maximum light occurs at the phase of minimum $V$-band light curve fitted from the Fourier expansion.}, denoted as $\Phi_{max}$, can be obtained from the fitted $V$-band light curves. Then, the light curves in {\it both} $V$- and $I$-band can be phase-shifted according to the following relations:

     \begin{eqnarray}
       \phi_k(\mathrm{new}) & = & 2 k \pi \Phi_{max} + \phi_k(\mathrm{old}),
     \end{eqnarray}
 
     \ni where $\phi_k(\mathrm{old})$ are from equation (2). Hence, the zeroth phase of the new light curves will correspond to the ($V$-band) maximum light. Since in general the Cepheid light curves are skew-symmetric with slow declination from maximum to minimum, the phase of minimum light is located roughly in between phases $\sim0.6$ and $\sim0.8$. The averages of the phase at minimum light, $\phi_{min}$, for the various samples are summarized in Table \ref{tabmin}.

     From the new light curves after the phase shift with equation (3), the colour at $i\mathrm{th}$ phase for a Cepheid is defined as $(V_i-I_i)$, where $i\in[0,1]$. The extinction corrected colour at $i\mathrm{th}$ phase is calculated with the subtraction of $\Delta R\times E(B-V)$ to $(V_i-I_i)$, where $\Delta R=R_V-R_I=1.28$ \citep{kan04,tam03,uda99}. The $V$-band amplitude for the Cepheids is just $V_{Amp}=V_{min}-V_{max}$. Therefore, together with the published periods, the multi-phase PC and AC relations can be constructed by performing a standard linear regression at the particular phase $i$ for a given set of Cepheids. 


     \begin{table}
       \centering
       \caption{The average values of the $\phi_{min}$ for the long ($\log P>1$), short ($\log P<1$) and all (long$+$short) period Cepheids.}
       \label{tabmin}
       \begin{tabular}{lccc}\hline
         Period-Range  & GAL & LMC & SMC  \\
         \hline \hline
         All   & 0.675 & 0.733 & 0.738 \\
         Long  & 0.641 & 0.671 & 0.644 \\ 
         Short & 0.687 & 0.740 & 0.754 \\
         \hline
       \end{tabular}
     \end{table}


     \begin{figure}
       \hbox{\hspace{0.1cm}\epsfxsize=7.5cm \epsfbox{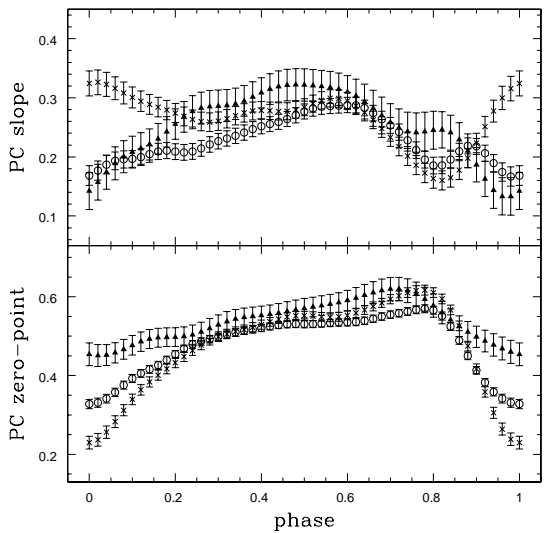}}
       \caption{The PC relations as function of phases for all (long$+$short) of the Cepheids in the samples. The symbols are: Galactic Cepheids = filled triangles; LMC Cepheids = open circles; and SMC Cepheids = crosses. Phase zero corresponds to the maximum light.}
      \label{figpcall}
     \end{figure}


\section{The Multi-Phase PC Relations}

     The slopes and the zero-points of the PC relations as function of pulsation phases ($\Phi$) for all of the Galactic, LMC and SMC Cepheids in our samples are plotted and compared in Figure \ref{figpcall}. Similar plots are presented in Figure \ref{figpclong} \& \ref{figpcshort} for the long and short period Cepheids, respectively. In these figures, phase zero ($\Phi=0$) corresponds to the maximum light. Our choice of the fiducial period $\log P = 1$ is governed by the work of Paper I, \citet{san04} and \citet{nge05}. Because of the large numbers of Cepheids, dividing the sample into two at this period does not significantly affect the value of the slope or zero-point at a given phase. That is, if the true underlying relation is consistent with two lines of differing slopes but continuous or nearly continuous at a period close to 10 days, then there are sufficient numbers of Cepheids on either side of 10 days to adequately represent this true relation \citep{nge05}. From these figures, we found that:

\begin{enumerate}
\item In Figure \ref{figpcall}, the general form of the PC relations as a function of phase in these three galaxies appear to be similar. The plots rise to a maximum at a phase between 0.5-0.6 (except for the SMC slopes). The LMC and SMC PC relations have a local minimum in the PC slope at a phase close to 0.8.

\item Long period Cepheids display their steepest PC slope at phases around $\sim0.7$ - $\sim0.8$ (at the same time the zero-points drops to the minimum values). This slope is the greatest in both of the Galaxy and LMC.

\item In contrast, short period Cepheids in the Galaxy, LMC and SMC display a flattening of the PC slope at $\Phi \sim 0.8$ (at the same time the zero-points approach to the maximum values), with some suggestion that the phase at which the slope is lowest gets later as the metallicity decreases. For the LMC short period Cepheids, this slope is very close to zero. 


     \begin{figure*}
       \vspace{0cm}
       \hbox{\hspace{0.2cm}\epsfxsize=8.5cm \epsfbox{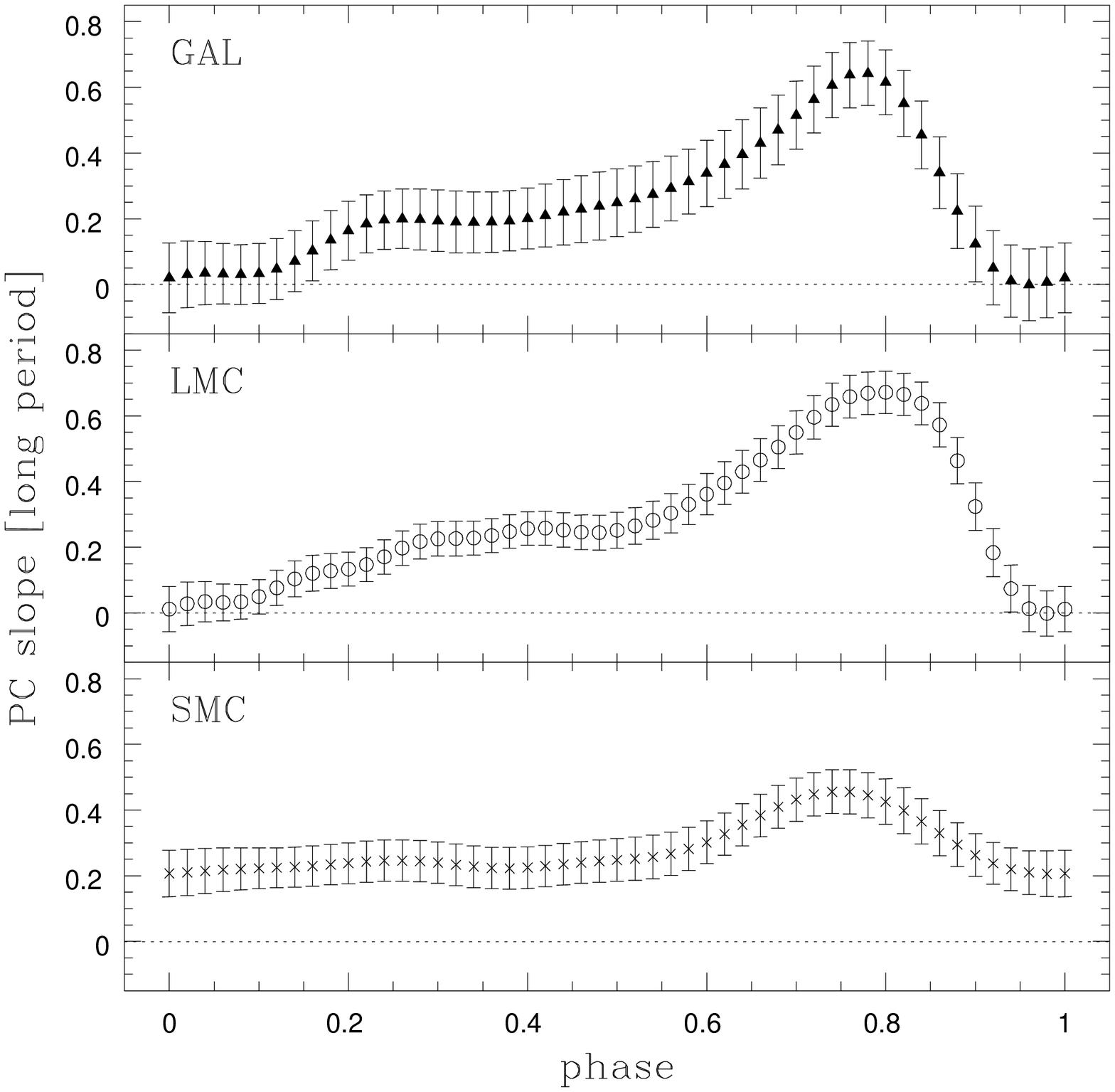}
         \epsfxsize=8.5cm \epsfbox{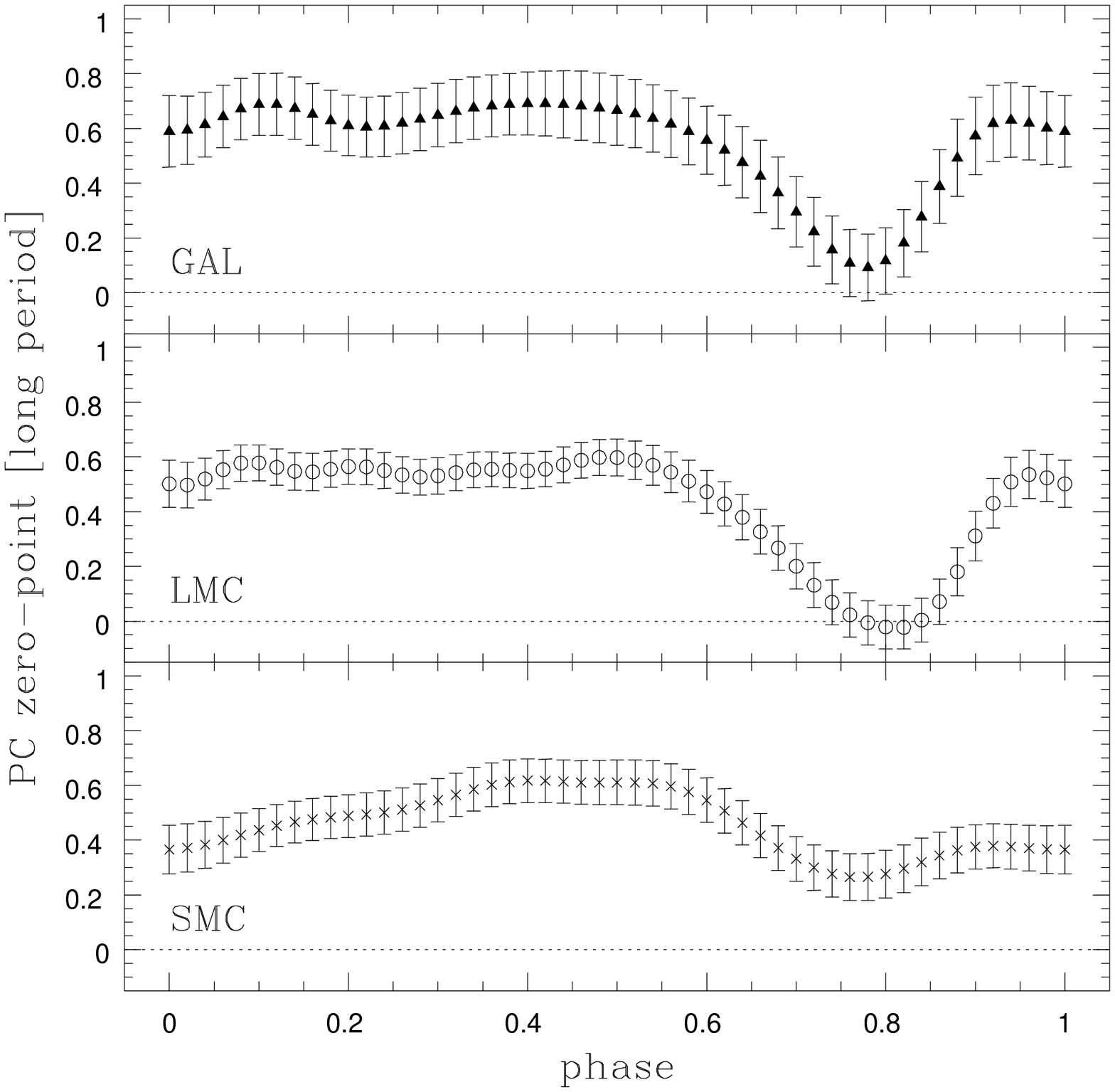}}
       \vspace{0cm}
       \caption{PC relation slope and zero-points as a function of phase for the long period ($\log P > 1$) Cepheids in the Galaxy, LMC and SMC. Phase zero corresponds to the maximum light. The dotted lines indicate the vanished slope/zero-point.}
       \label{figpclong}
     \end{figure*}


     \begin{figure*}
       \vspace{0cm}
       \hbox{\hspace{0.2cm}\epsfxsize=8.5cm \epsfbox{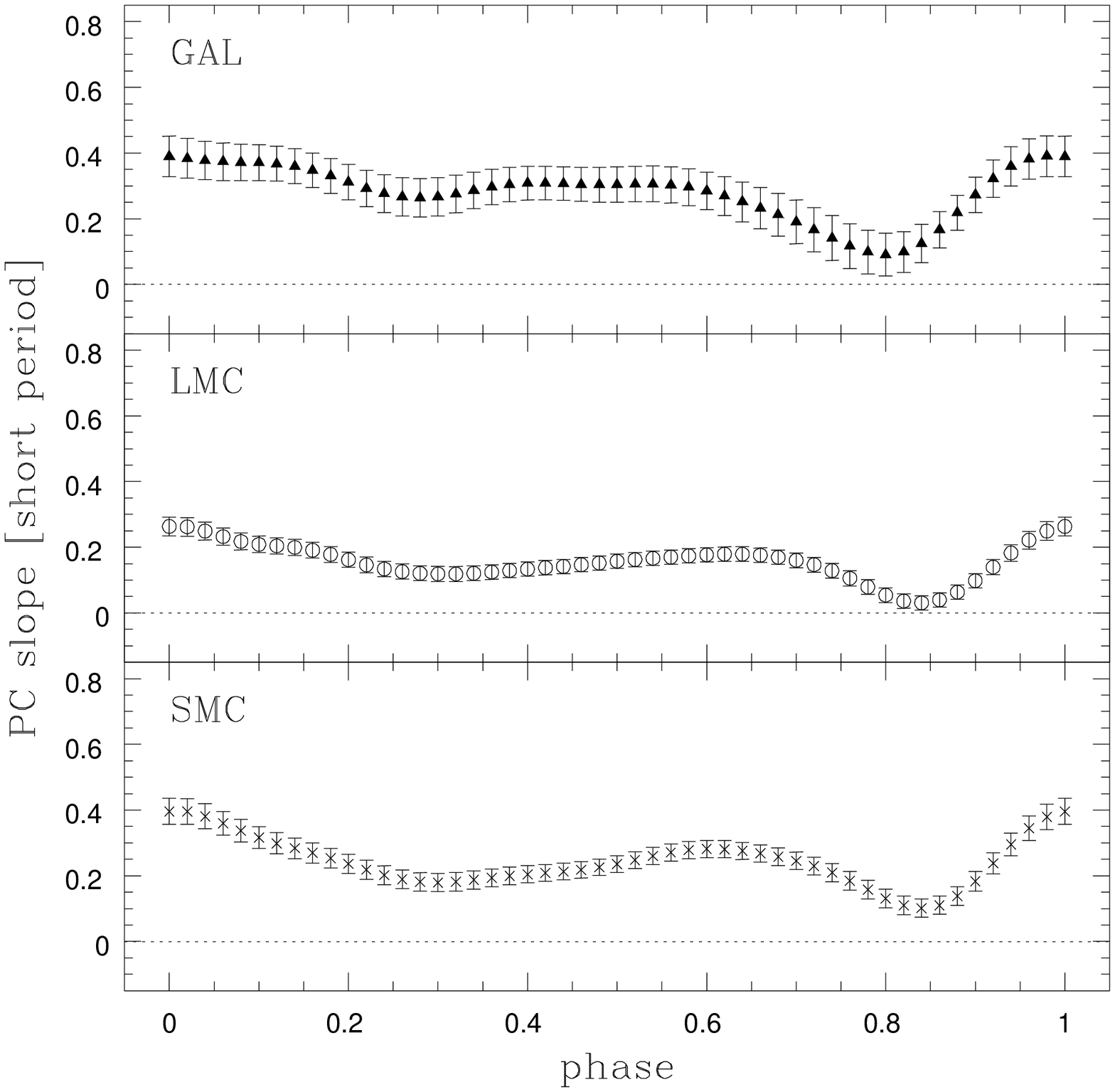}
         \epsfxsize=8.5cm \epsfbox{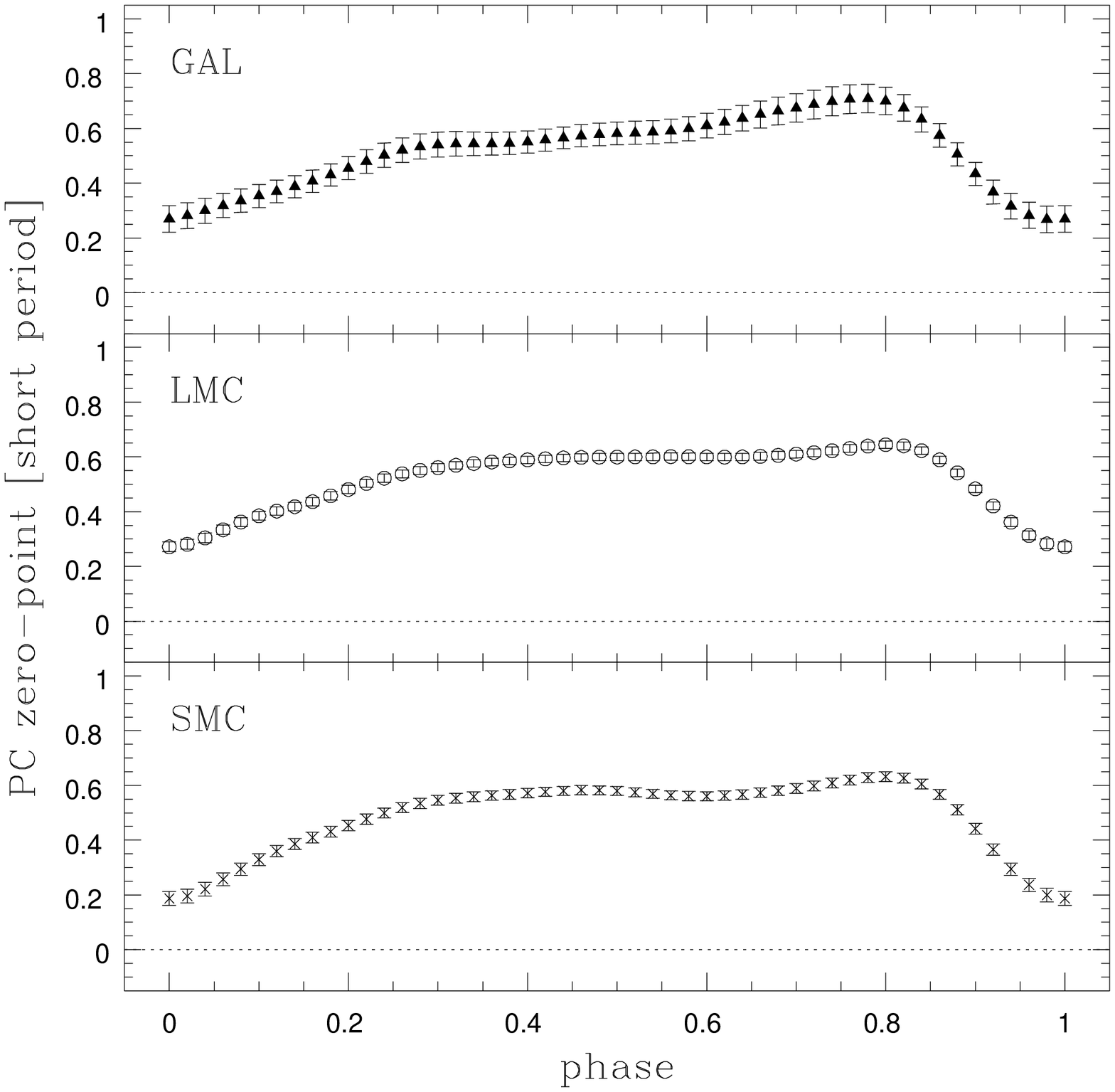}}
       \vspace{0cm}
       \caption{Same as Figure \ref{figpcshort}, but for the short period ($\log P < 1$) Cepheids.}
       \label{figpcshort}
     \end{figure*}

\item From Figure \ref{figpclong}, both LMC and Galactic long period Cepheids have a zero PC relation slope at phases close to maximum light. For the Galactic long period Cepheids, the slope is nearly zero between the phases $\sim 0.90$-$\sim0.13$ whereas this range is reduced in the LMC $\sim 0.95$-$\sim 0.08$. The long period SMC Cepheids have a non-zero slope in the PC plots for phases close to maximum light. In fact, the flatness of the Galactic PC relation at maximum light (at phase zero) starts at a period of $\log P \gtrsim 0.8$, as suggested in Paper II. This is confirmed by calculating the PC relation at maximum light for Galactic Cepheids with $\log P > 0.8$ (where the slope is $0.04\pm0.05$) and from the top-left panel of Figure \ref{pcvar}. However, this happens at $\log P \sim 1.0$ for the LMC Cepheids.

\end{enumerate}

\ni One important feature from these figures is that the slopes of the PC relations approach zero and maximum values for the short and long period Cepheids, respectively, at $\Phi \sim0.8$. Note that some interesting behavior around phase of $\sim0.8$ has been reported in the literature. For example, the angular diameters as a function of phase display a bump or large deviation from the normal fit around this phase \citep{fou03,sto04,gie05}. This is probably due to the presence of shock near minimum light \citep{fou03}. \citet{but93} and \citet{but96} have also reported some interesting hydrodynamics effects at phase around 0.8 from spectroscopic lines observations for a few Galactic Cepheids. However, \citet{lan95} demonstrated that this phase may not be problematic for the Baade-Wesselink type applications.

\subsection{Comparisons of the Galactic and LMC PC relations}

     As mentioned before, the mean light PC relation is linear for the Galactic Cepheids but non-linear for the LMC Cepheids, even thought the PC relations as a function of phase for both galaxies are similar to each others (especially for the long period PC relations) rather than to the SMC PC relations (see Figure \ref{figpclong} \& \ref{figpcshort}). Therefore it is interesting to compare the Galactic and the LMC PC relations at various phases to investigate this. In what follows we concentrate on the differences between the Galaxy and LMC PC relations. 

\begin{figure}
       \hbox{\hspace{0.1cm}\epsfxsize=7.5cm \epsfbox{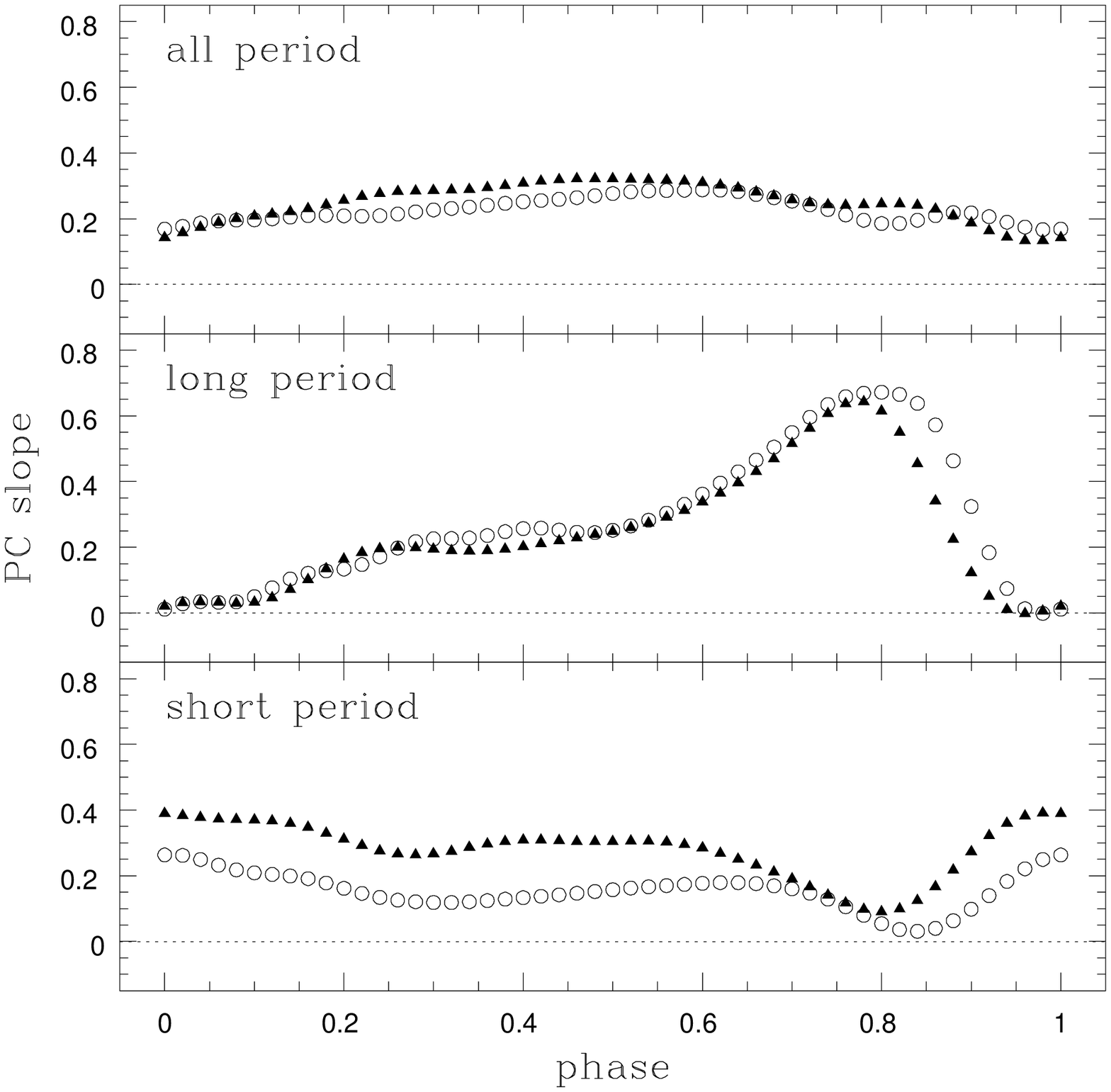}}
       \caption{Variations of the slopes for the empirical Galactic and LMC PC relations as a function of pulsational phase. The solid triangles and the open circles are for the Galactic and LMC PC relations, respectively. The dotted lines represent the zero slope, i,e., the slopes near these lines will have a flat PC relation. The maximum light occurs at the zeroth phase. Error bars are omitted for clarity.}
      \label{figcom_p}
     \end{figure}

\begin{figure}
       \hbox{\hspace{0.1cm}\epsfxsize=7.5cm \epsfbox{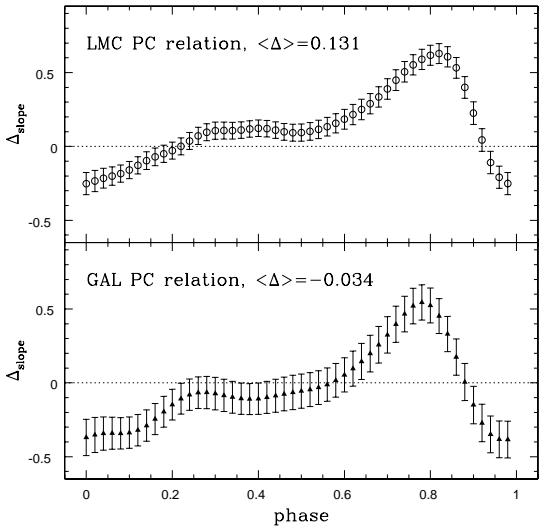}}
       \caption{Difference of the slopes for the long and short period PC relation as a function of phase, for the LMC (top panel) and the Galactic (bottom panel) Cepheids. The dotted lines represent the zero difference.}
      \label{figdiff}
     \end{figure}

     For ease of comparison, Figure \ref{figcom_p} combines the Galactic and LMC results in a single three panel plot with solid triangles and open circles representing Galactic and LMC Cepheids respectively. This plot provides evidence that the slopes of the long \& short period LMC PC relations are generally steeper \& shallower, respectively, than the Galactic counterparts at most of the phases. The difference in the slopes for long and short period PC relations ($\Delta_{\mathrm{slope}} = \mathrm{slope}_L-\mathrm{slope}_S$) is then plotted as a function of pulsational phase in Figure \ref{figdiff}. Even though $\Delta_{\mathrm{slope}}$ appears to be similar for the LMC and the Galactic PC relations\footnote{Besides the similar shapes of $\Delta_{\mathrm{slope}}$, the amplitudes of the differences for the Galactic and LMC PC relations are also similar, with amplitude, $|\Delta_{\mathrm{slope}}(MAX)-\Delta_{\mathrm{slope}}(MIN)|$, of $\sim0.9$.}, a careful examination of Figure \ref{figdiff} reveals that, aside from differences near phase zero, that $\Delta_{\mathrm{slope}}$ of the Galactic PC relations is different from the zero for only a small fraction (about $\sim0.15$ at $\Phi\sim0.8$) of phase. In contrast, the fraction of the phases that $\Delta_{\mathrm{slope}}$ is different from $\Delta_{\mathrm{slope}}=0$ for the LMC PC relations is $\sim0.6$: much larger than the Galactic value. This results in mean differences of $-0.034$ and $0.131$ for the Galactic and LMC PC relations, respectively. Together with the results from Figure \ref{figcom_p}, this is what leads to a non-linear LMC PC(mean) relation and a linear Galactic PC(mean) relation.


     \begin{figure}
       \hbox{\hspace{0.1cm}\epsfxsize=7.5cm \epsfbox{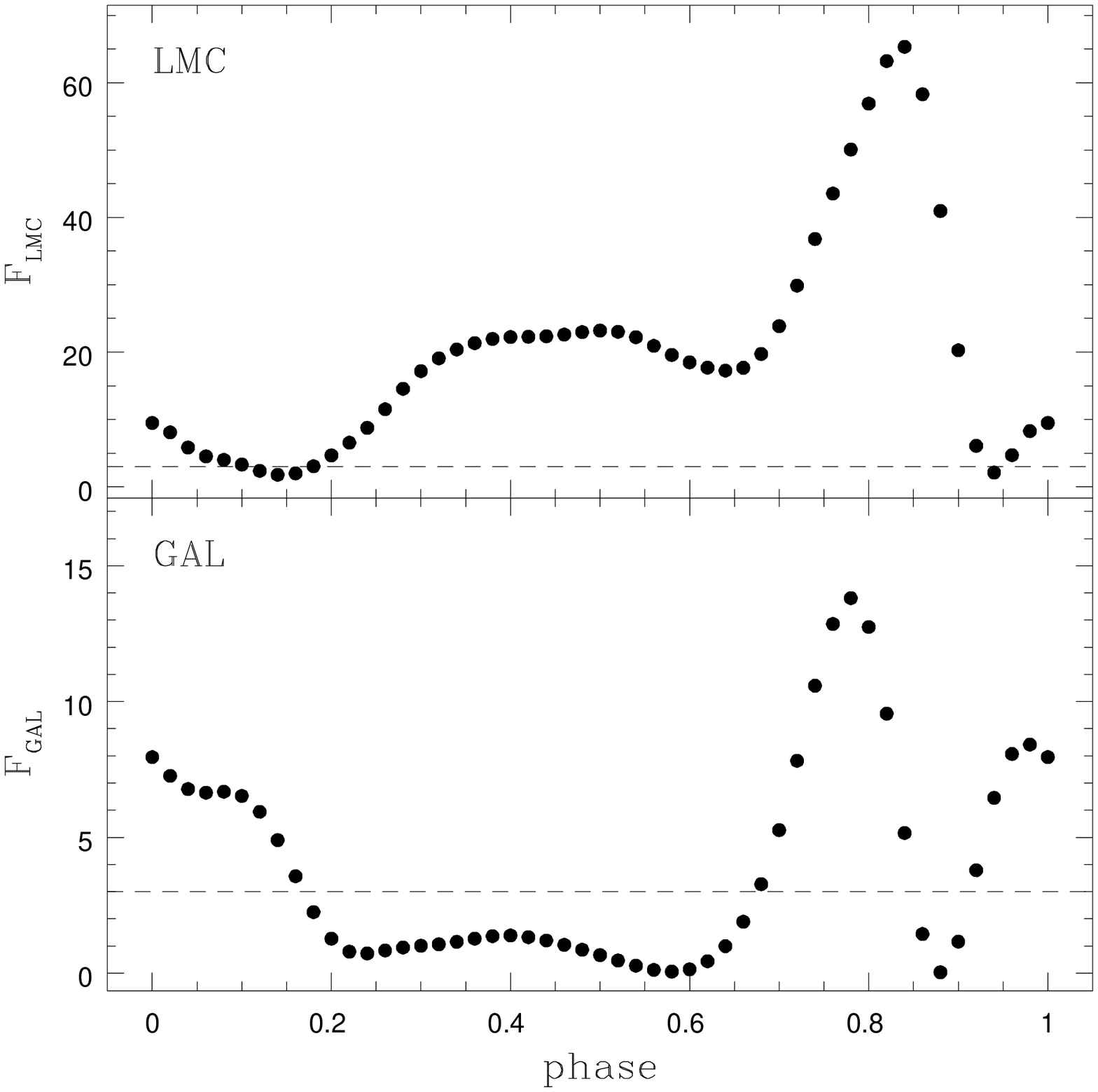}}
       \caption{Comparison of the $F$-values for the PC relation as a function of phase for the LMC (upper panel) and the Galactic (lower panel) Cepheids. The dashed lines represent the 95\% confidence level (or $p[F]=0.05$), corresponding to $F\sim3.0$. Hence, the $F$-values above the dashed line indicate that the PC relation is non-linear, while the $F$-values below this line suggest the PC relation is linear.}
      \label{figcom_f}
     \end{figure}

     The $F$-tests (see Paper I for details) are then applied to the Galactic and LMC multi-phase PC relations. The plots of the $F$-values as a function of phase are presented in Figure \ref{figcom_f}. In this figure, the dashed lines represent the 95\% confidence level (or $p[F]=0.05$), corresponding to $F\sim3.0$. Therefore, the $F$-values above the dashed line indicate that the PC relation is non-linear, while the $F$-values below this line indicate the PC relation is linear. From this figure, it can be seen that the LMC PC relations are not linear for most of the pulsational phases, except around $\Phi\sim0.15$ and $\Phi\sim0.95$. In contrast, the Galactic PC relations are linear for a large fraction of the pulsational phases, from $\Phi\sim0.15$ to $\Phi\sim0.70$ and again at $\Phi\sim0.88$. Therefore, these properties make the LMC PC(mean) and Galactic PC(mean) relation to be non-linear and linear respectively. Although the non-linearity of the Galactic and LMC PC relations at maximum light (with large $F$-values) has been mentioned previously, Figure \ref{figcom_f} shows that this non-linearity extends in pulsational phases around the maximum light in both galaxies. Further, the non-linear PC relations around $\Phi\sim0.8$ (the huge peak in Figure \ref{figcom_f}) are seen not only in the LMC Cepheids but also for the Galactic Cepheids. This is because the slopes for both Galactic and LMC long and short period Cepheids reach a maximum and minimum, respective, at this phase (the zero-points also display the opposite trends). Even though both Galactic and LMC PC relations are not linear around $\Phi \sim0.8$, Figure \ref{figcom_f} shows why the mean light Galactic and LMC PC relations are linear and non-linear respectively, because the phase that corresponds to the mean light is located in between $\Phi \sim0.4$ and $\Phi \sim0.6$.

\begin{figure*}
       \centering
       \hbox{\hspace{0.25cm}\epsfxsize=7.1cm \epsfbox{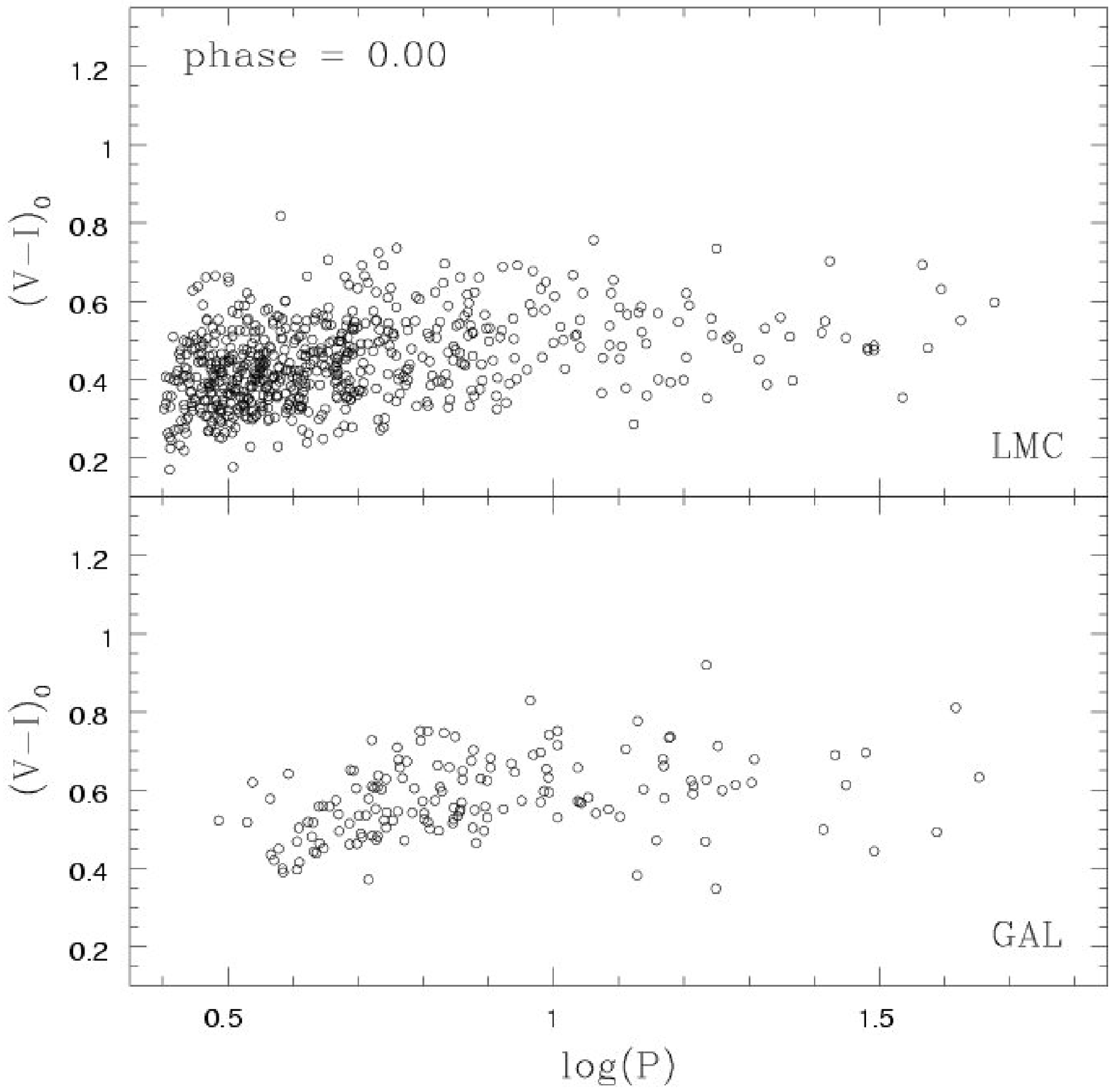}
         \epsfxsize=7.1cm \epsfbox{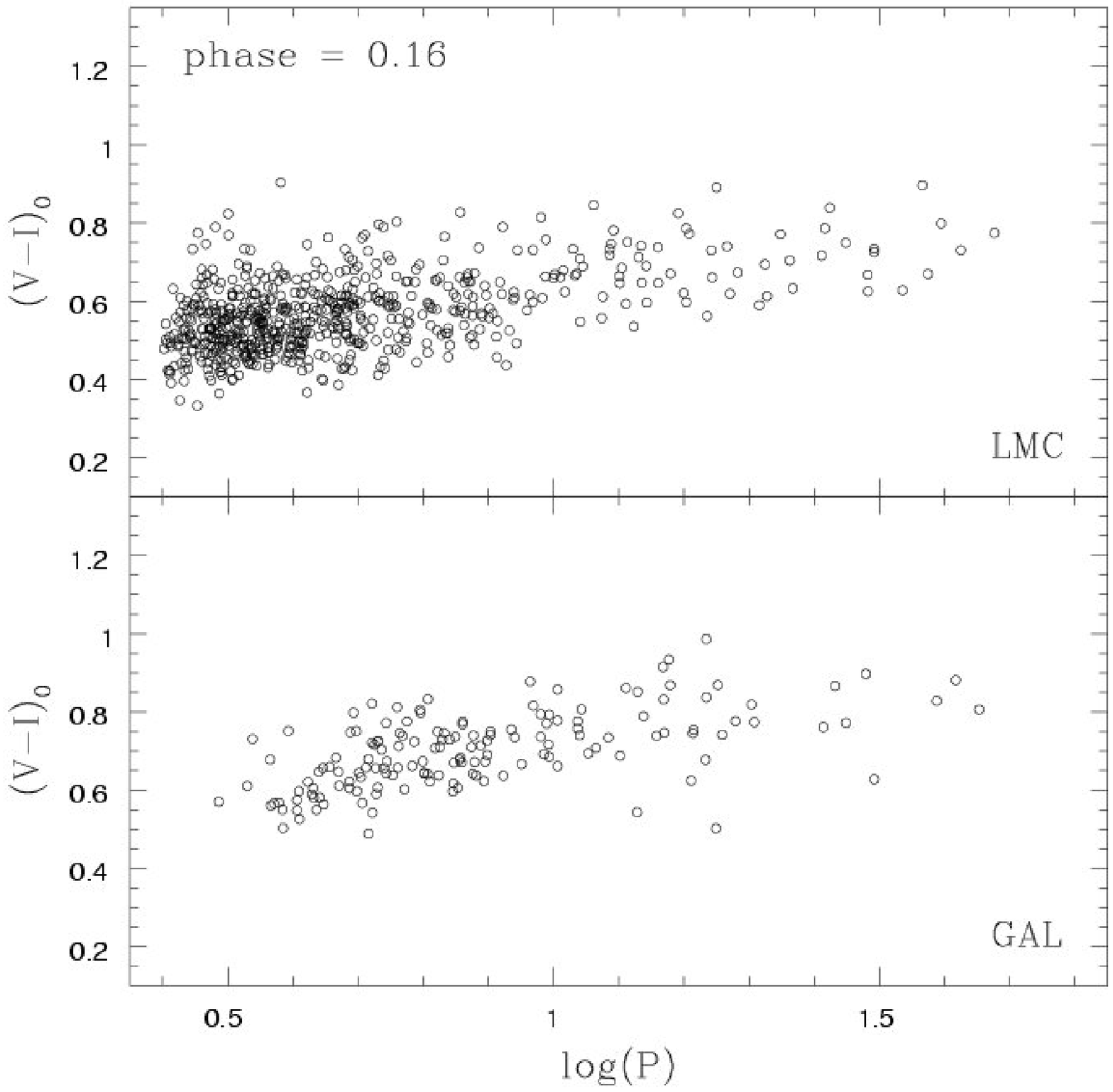}}
       \hbox{\hspace{0.25cm}\epsfxsize=7.1cm \epsfbox{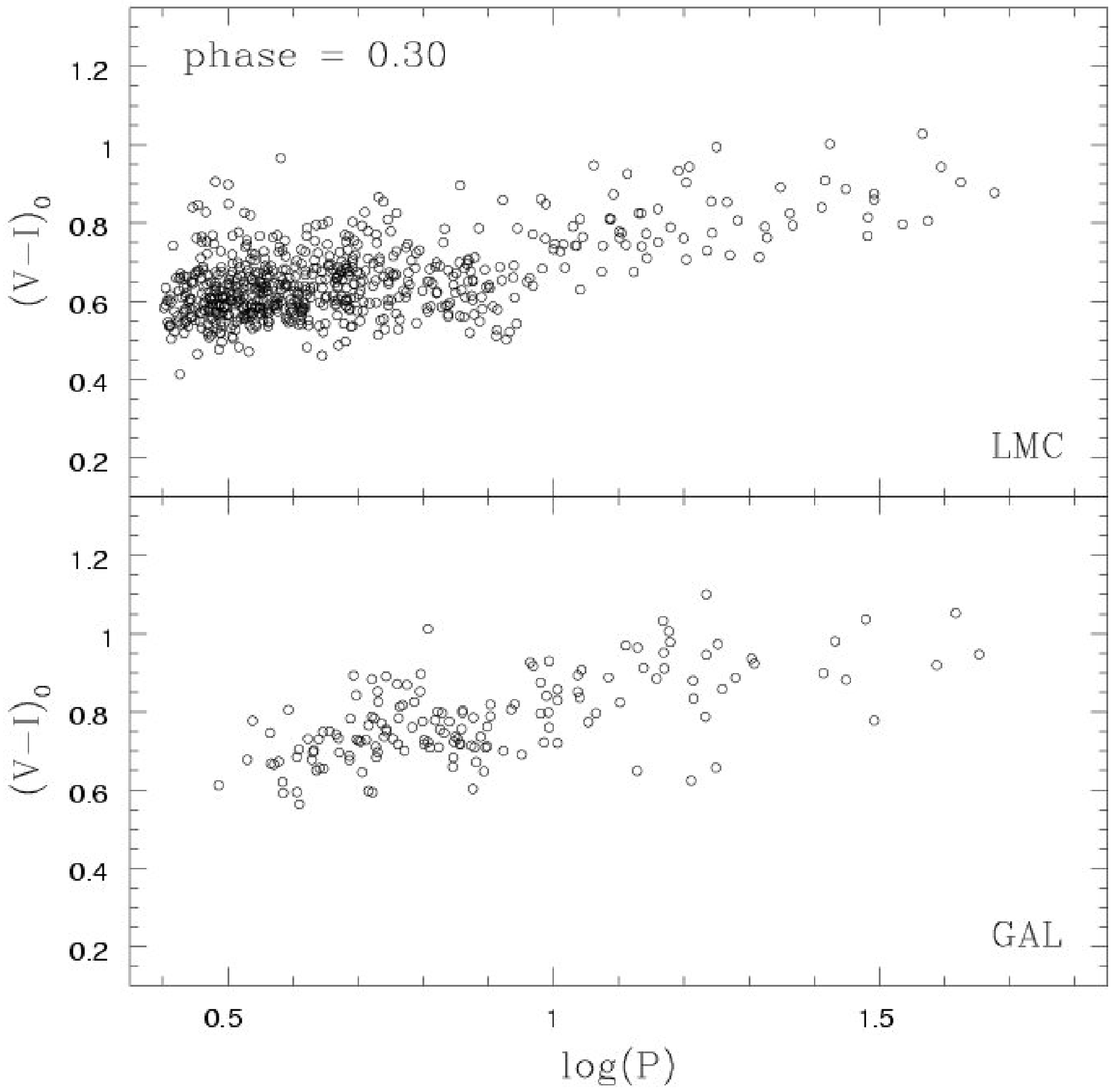}
         \epsfxsize=7.1cm \epsfbox{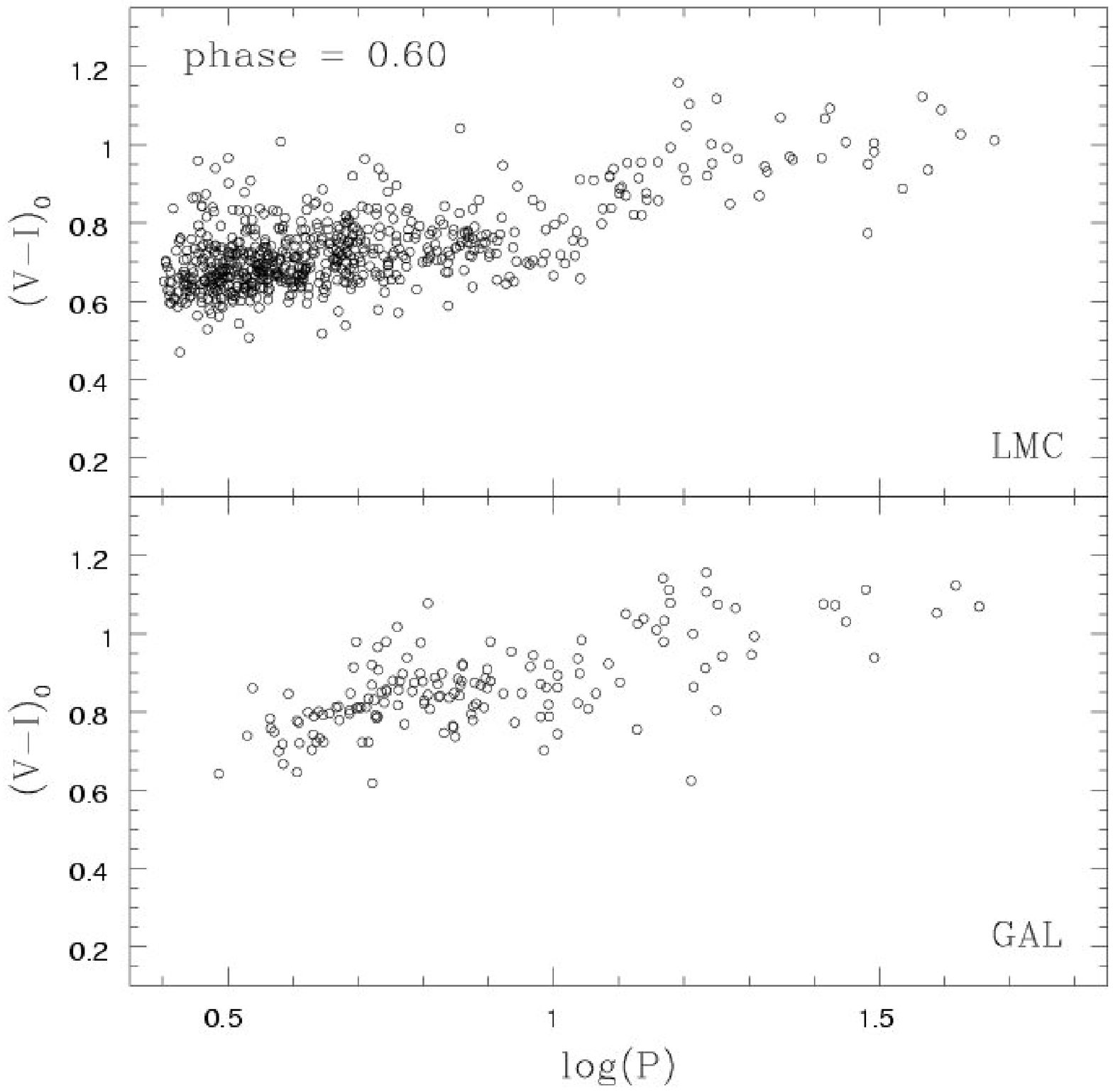}}
       \hbox{\hspace{0.25cm}\epsfxsize=7.1cm \epsfbox{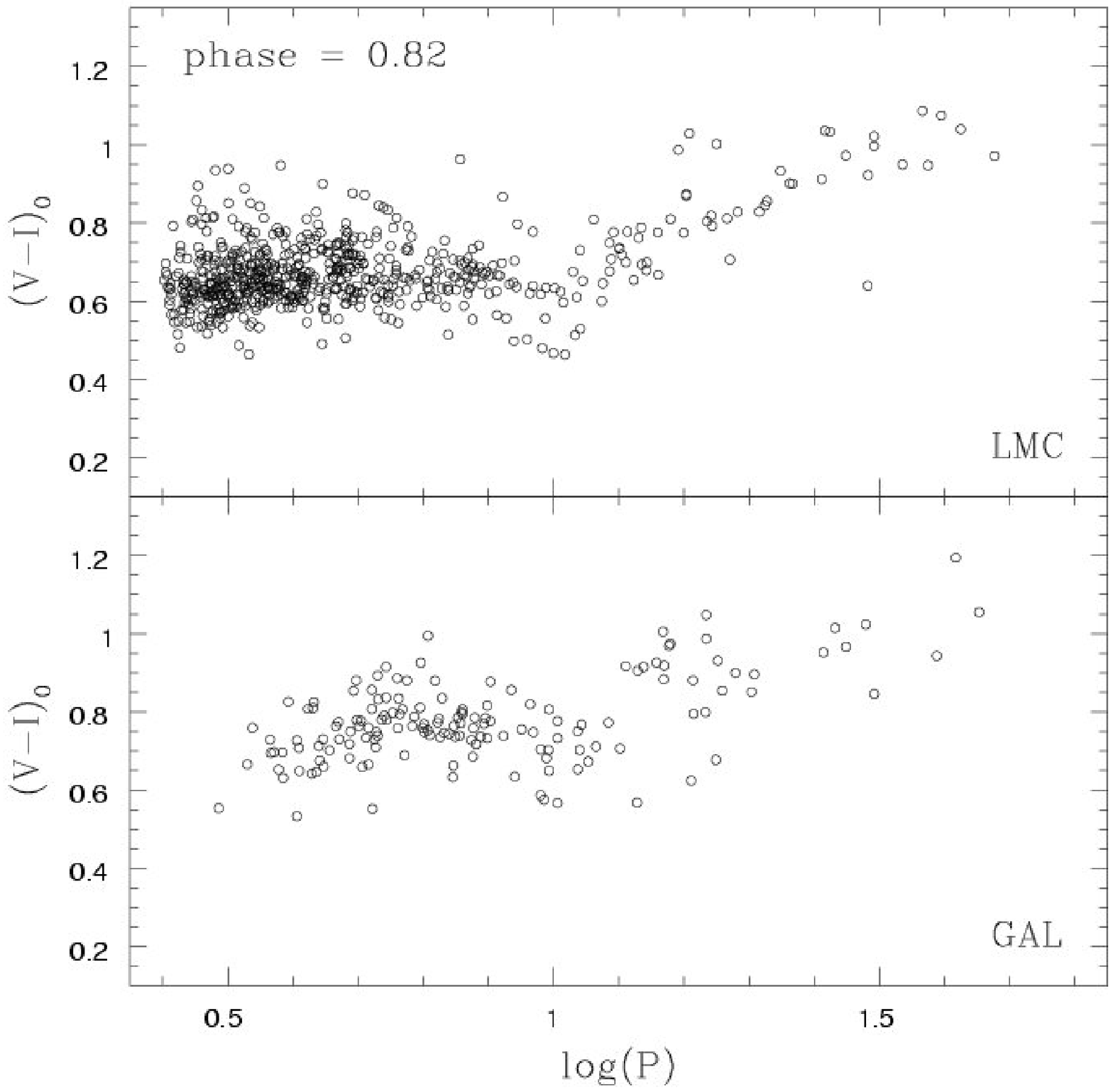}
         \epsfxsize=7.1cm \epsfbox{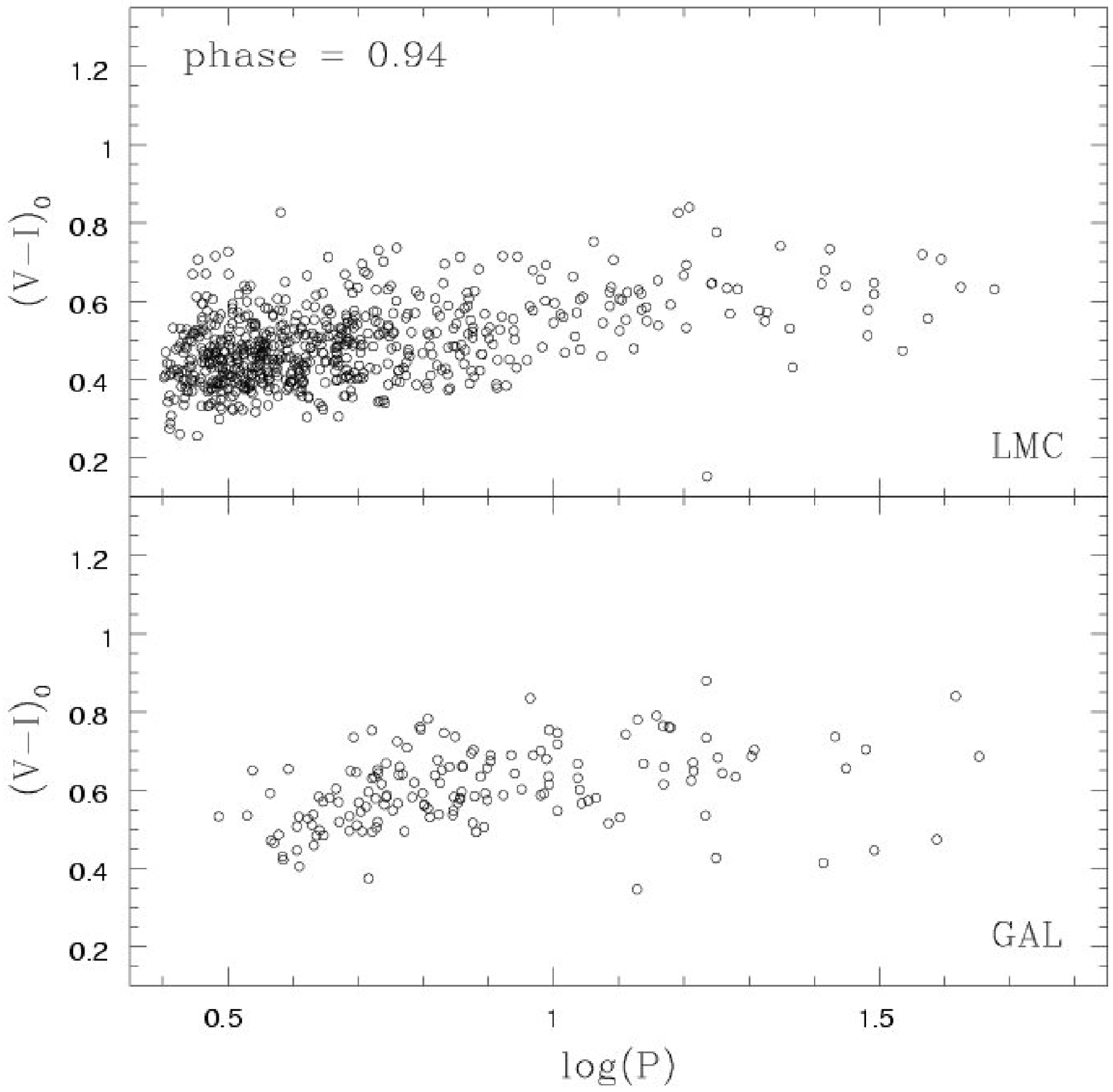}}
       \vspace{0cm}
       \caption{Comparisons of the PC relations for the LMC and Galactic Cepheids at various phases.}
       \label{pcvar}
       \end{figure*}

     Figure \ref{pcvar} presents snapshots of the Galactic and LMC PC relations at certain phases beginning with maximum light (phase zero). At maximum light, both of the LMC and Galactic Cepheids follow flat PC relations for Cepheids with $\log P \gtrsim 1.0$ and $\log P \gtrsim 0.8$, respectively. After maximum light, the LMC PC relation becomes linear near $\Phi \sim 0.15$ but the Galactic PC relation stays non-linear around the same phase (see Figure \ref{figcom_f}). This is portrayed in the upper-right panels of Figure \ref{pcvar} for $\Phi=0.16$. Then the LMC \& Galactic PC relations become non-linear \& linear, respectively, for a large fraction of phases ($\Phi \sim0.2$ - $\sim 0.7$), as portrayed in Figure \ref{figcom_f} and middle panels of Figure \ref{pcvar}. Phases around 0.8 are interesting because the PC relations for both of the Galactic and LMC Cepheids are non-linear with very large $F$-values (the peaks in Figure \ref{figcom_f}). Hence the PC relations are plotted in the lower-left panel of Figure \ref{pcvar} for $\Phi=0.82$. The non-linearity of the PC relations at this phase (in both galaxies) with a sharp break at $\log P\sim 1$ is clearly evident. Furthermore, the slopes of the short period PC relations approach zero around this phase (LMC: slope$=0.031\pm0.022$ at $\Phi\sim0.84$; Galactic: slope$=0.091\pm0.065$ at $\Phi \sim0.80$). However, the plots of the PC relation at $\Phi=0.82$ in Figure \ref{pcvar} show that the short period LMC PC relation is indeed nearly flat around this phase, but the short period Galactic PC relation can be broken further into two PC relations with a break at $\log P \sim0.8$. The short period Galactic PC relation has a positive slope for $\log P<0.8$ and a negative slope for $0.8<\log P <1.0$, hence the overall slope is close to zero as seen in the lower panel of Figure \ref{figcom_p}. Finally, PC relations at $\Phi=0.94$ are plotted in the lower-right panel of Figure \ref{pcvar} to show the transition from $\Phi\sim0.8$ to maximum light.

 
\subsection{The SMC PC Relations}

 
     \begin{figure*}
       \centering
       \hbox{\hspace{0.25cm}\epsfxsize=7.1cm \epsfbox{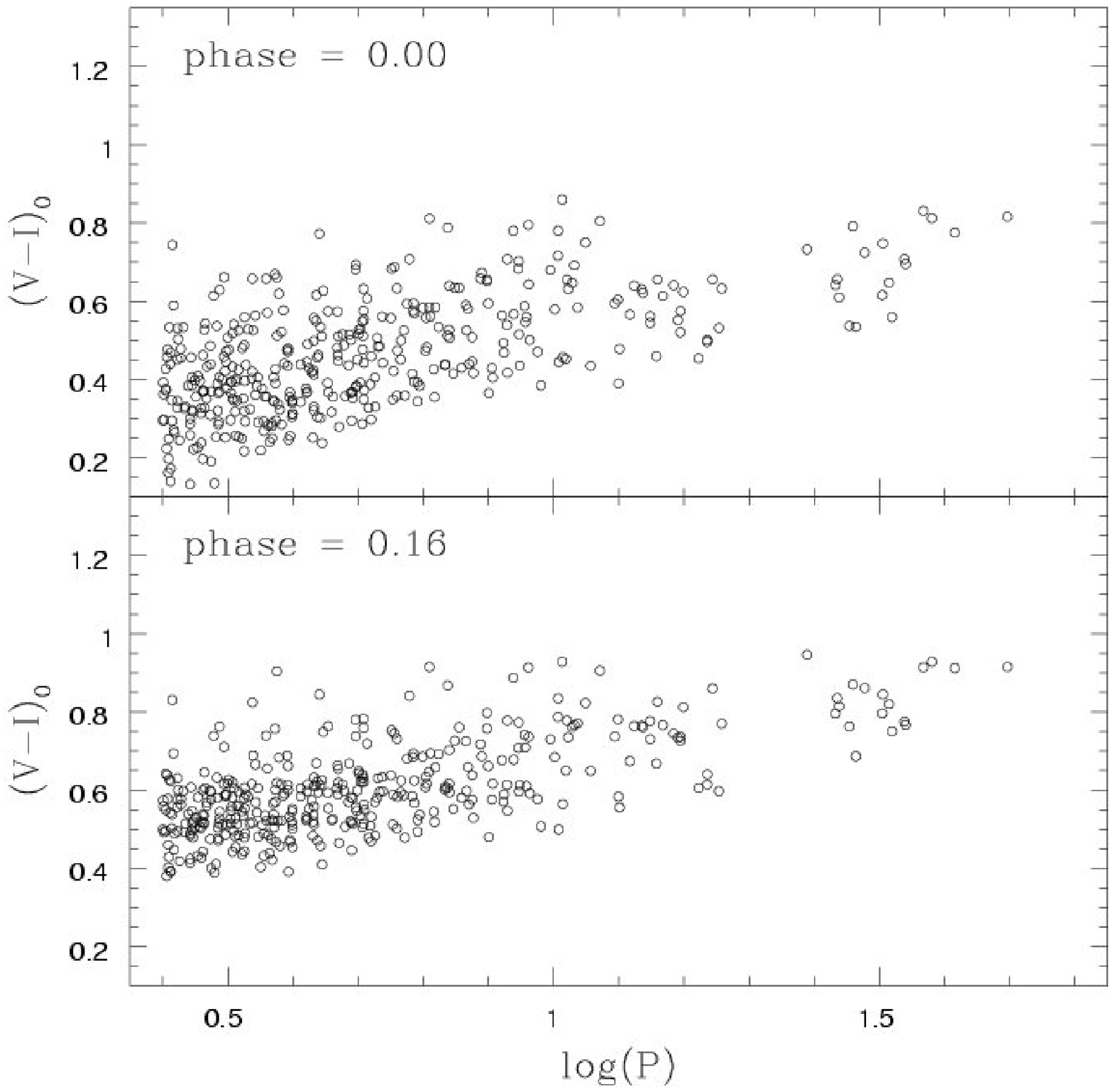}
         \epsfxsize=7.1cm \epsfbox{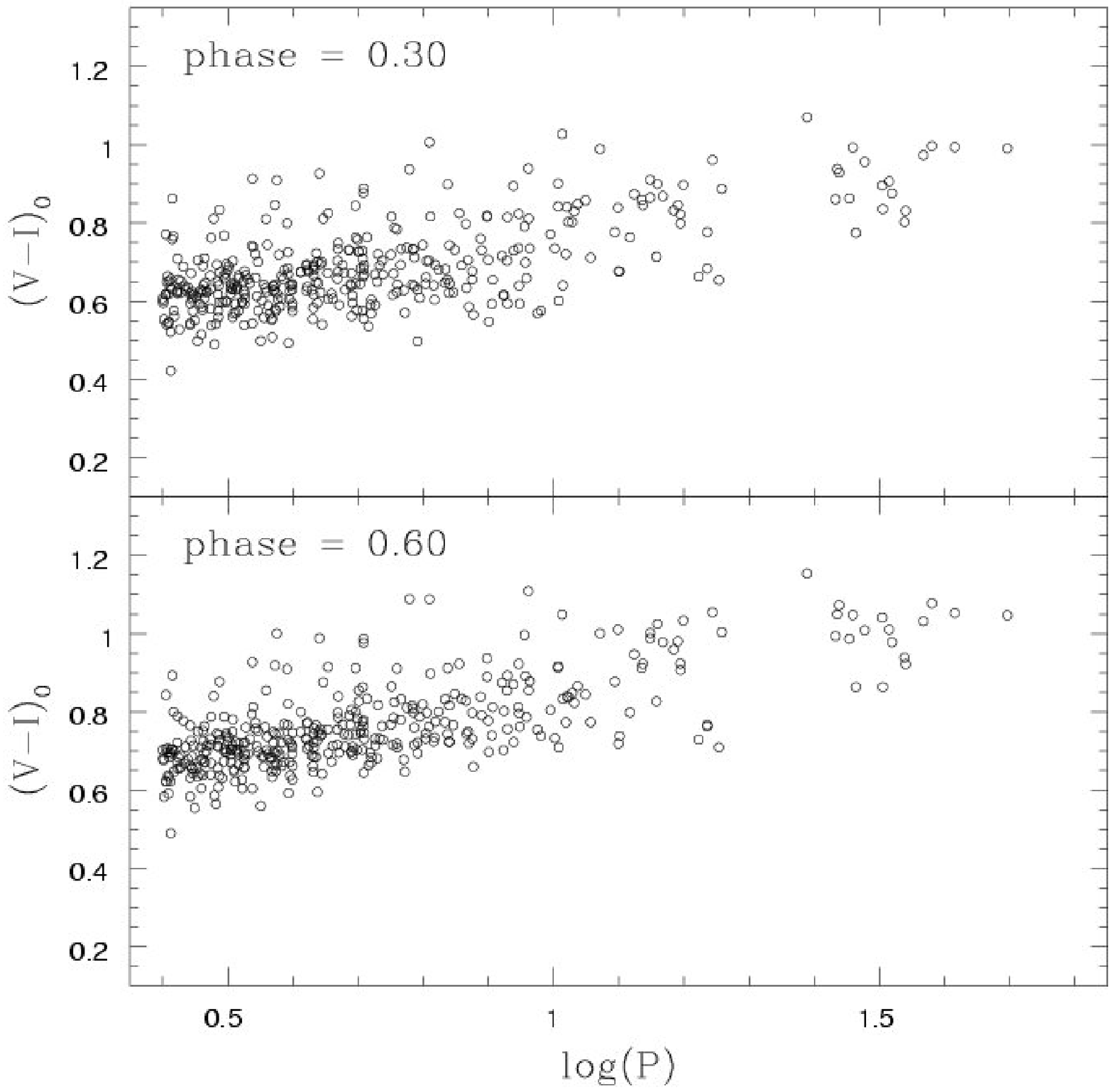}}
       \hbox{\hspace{0.25cm}\epsfxsize=7.1cm \epsfbox{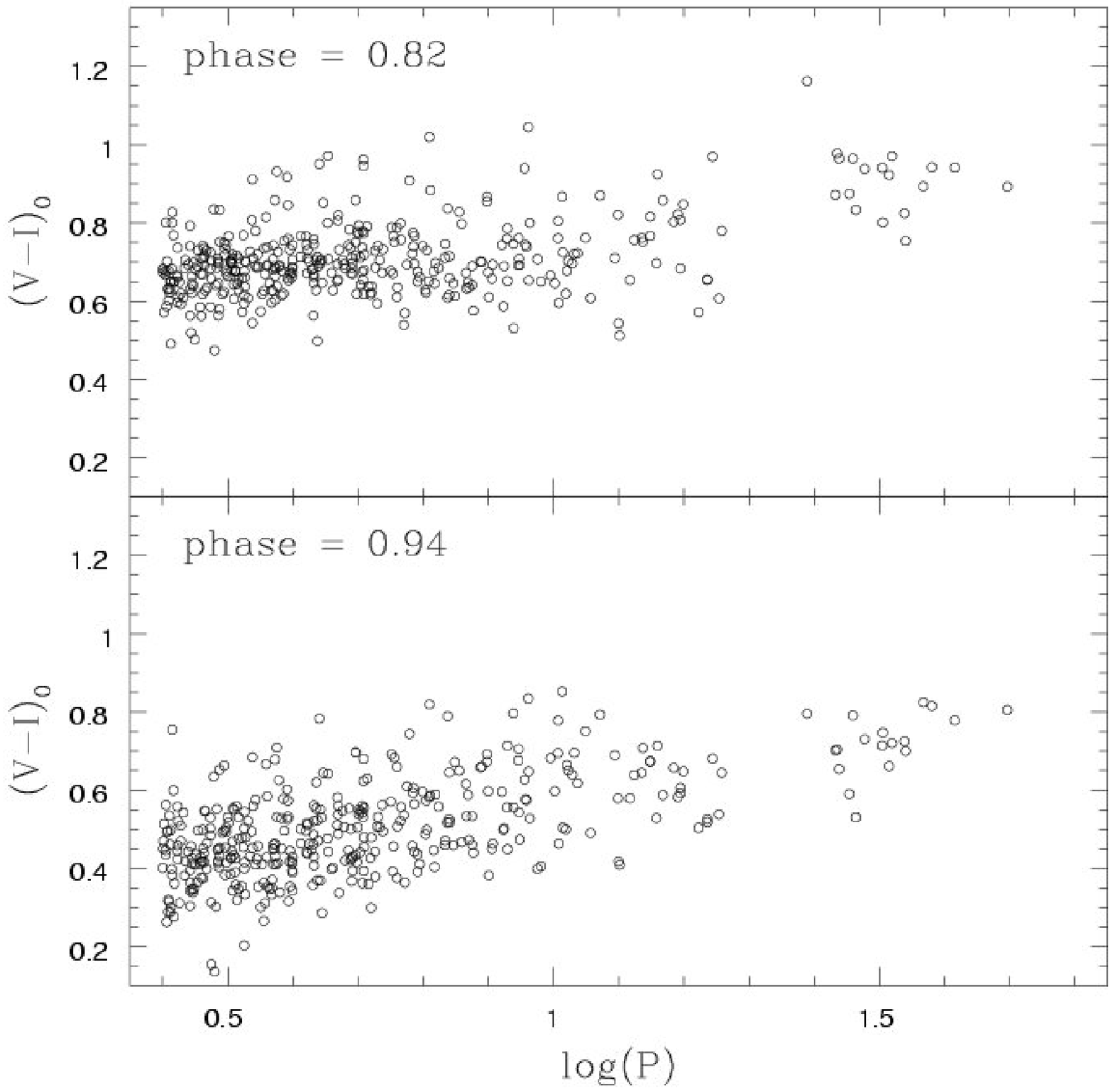}
         \epsfxsize=7.1cm \epsfbox{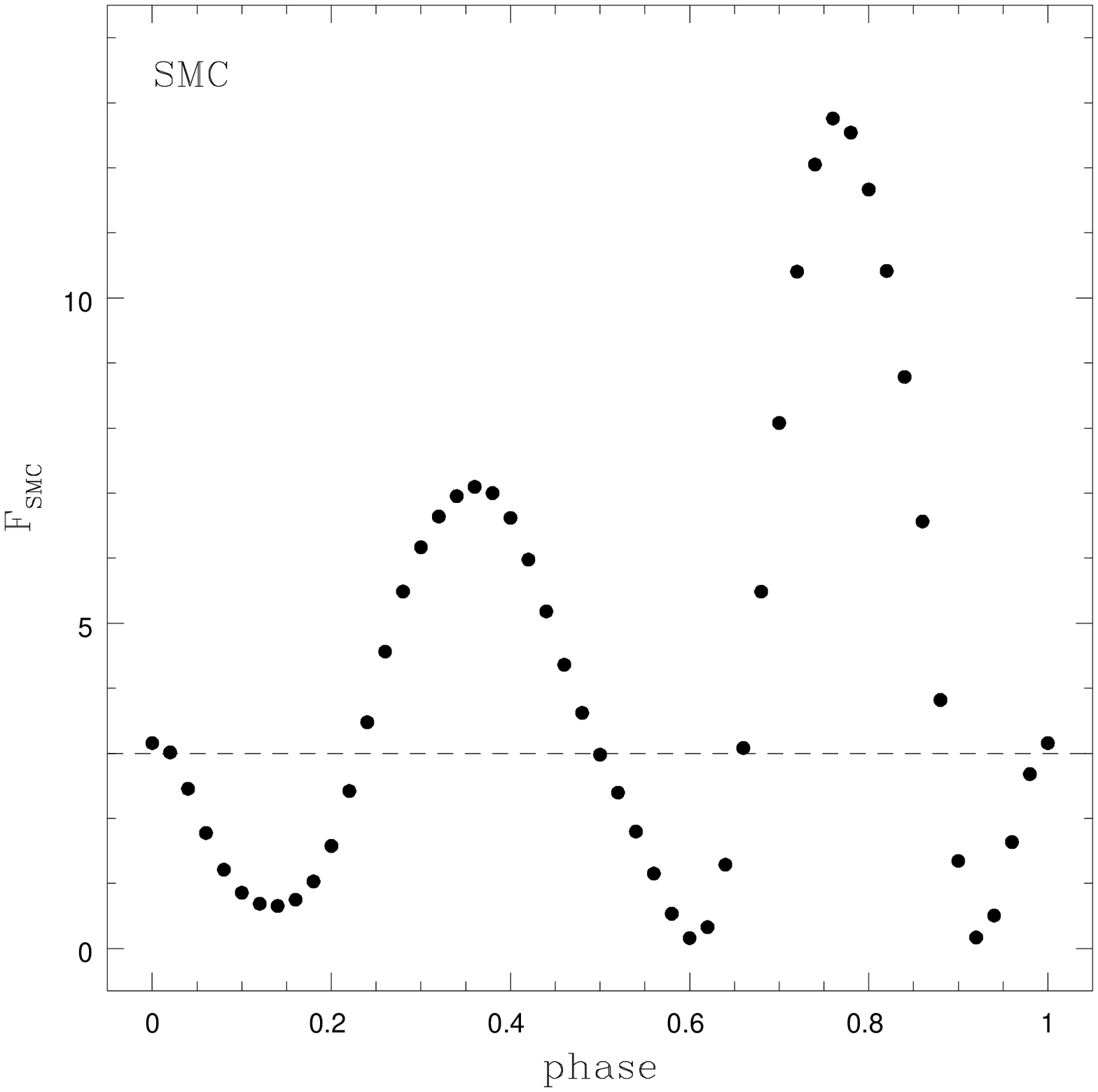}}
        \vspace{0cm}
       \caption{PC relations for SMC Cepheids at various phases that are the same as in Figure \ref{pcvar} and the plot of $F$-values as a function of phase.}
       \label{figsmc}
       \end{figure*}

For completeness, we include the snapshots for the SMC multi-phase PC relations (same layout as in Figure \ref{pcvar}) and the plot of the $F$-values as function of phase in Figure \ref{figsmc}. Compared to Figure \ref{figcom_f}, there are some similarities and differences for the behavior of non-linearity of the SMC PC relations as a function of phase to the Galactic and LMC counterparts. For example, the non-linear PC relation at maximum light is only marginal for the SMC Cepheids, in contrast to the Galactic and LMC PC(max) relations. Between $\Phi \sim0.2$ and $\Phi \sim0.6$, the Galactic and LMC PC relations are linear and non-linear, respectively. However, the SMC PC relations switch from linear around $\Phi \sim0.25$ to become non-linear then back to linear near $\Phi \sim0.50$. This ``secondary peak'' is absent for the Galactic and LMC PC relations. For SMC Cepheids, the phases that correspond to the mean light are around $\Phi \sim0.5$ and $\Phi \sim0.7$: hence the mean light PC relation is linear. In term of similarity, the significance of non-linear PC relations around $\Phi \sim0.8$ for SMC Cepheids is also clearly evident, which is also seen in the Galactic and LMC PC relations. The non-linearity of the PC relation around $\Phi \sim0.8$ seems to be a universal feature for the classical Cepheids.


\section{The Multi-Phase AC Relations}

\begin{figure}
       \hbox{\hspace{0.1cm}\epsfxsize=7.5cm \epsfbox{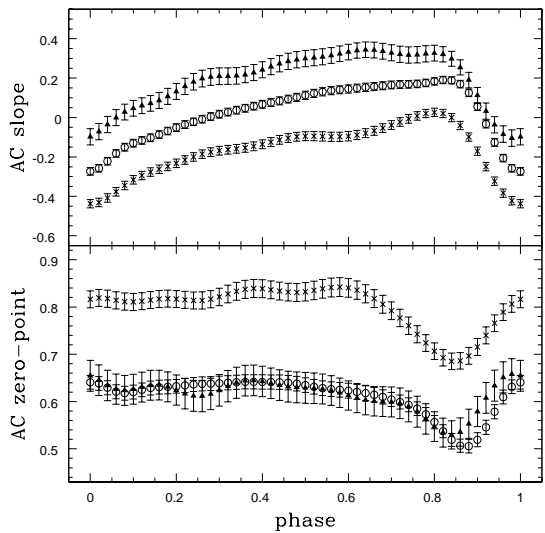}}
       \caption{Same as Figure \ref{figpcall}, but for the AC relations.}
      \label{figacall}
     \end{figure}

\begin{figure*}
       \vspace{0cm}
       \hbox{\hspace{0.2cm}\epsfxsize=8.5cm \epsfbox{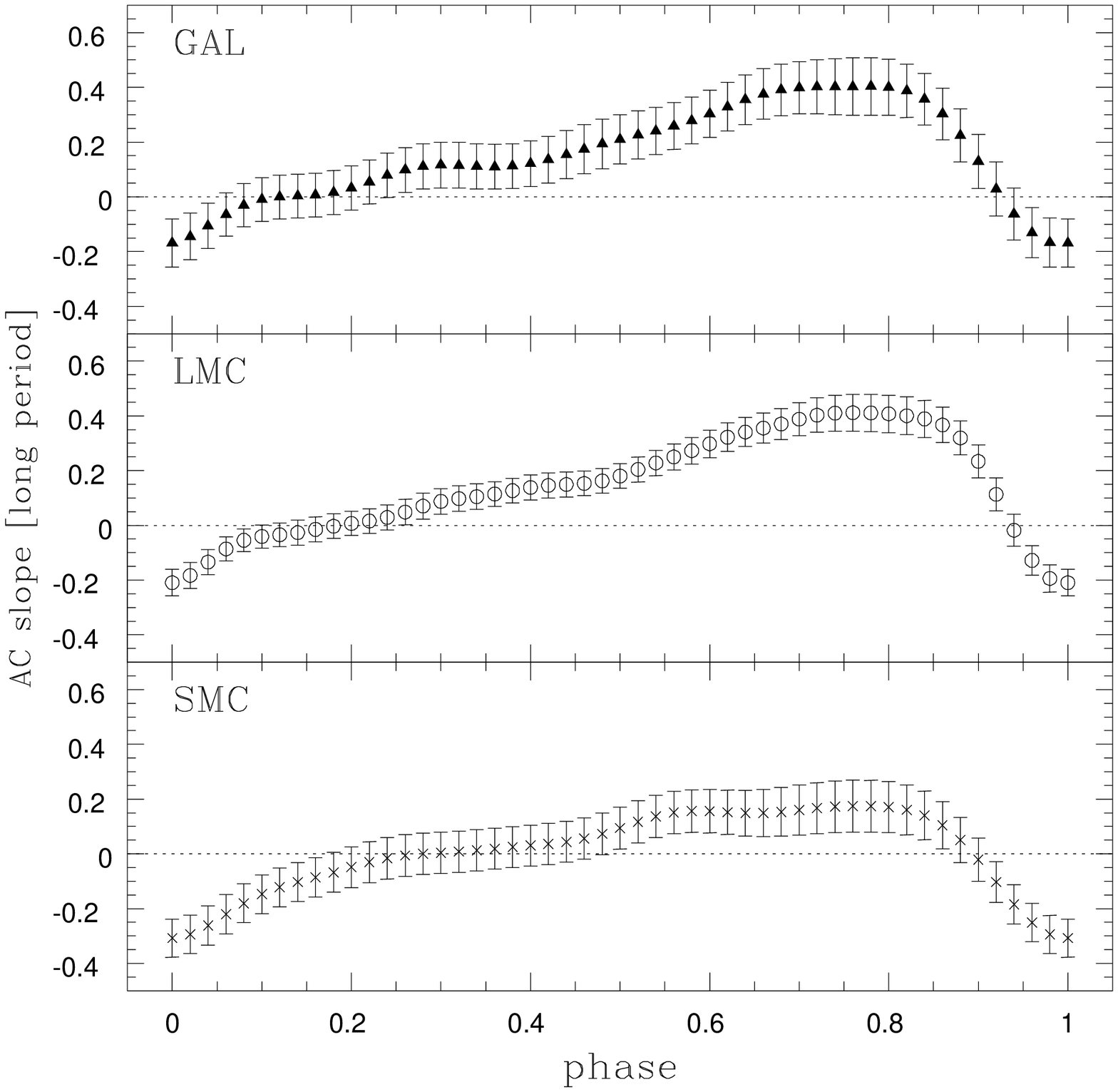}
         \epsfxsize=8.5cm \epsfbox{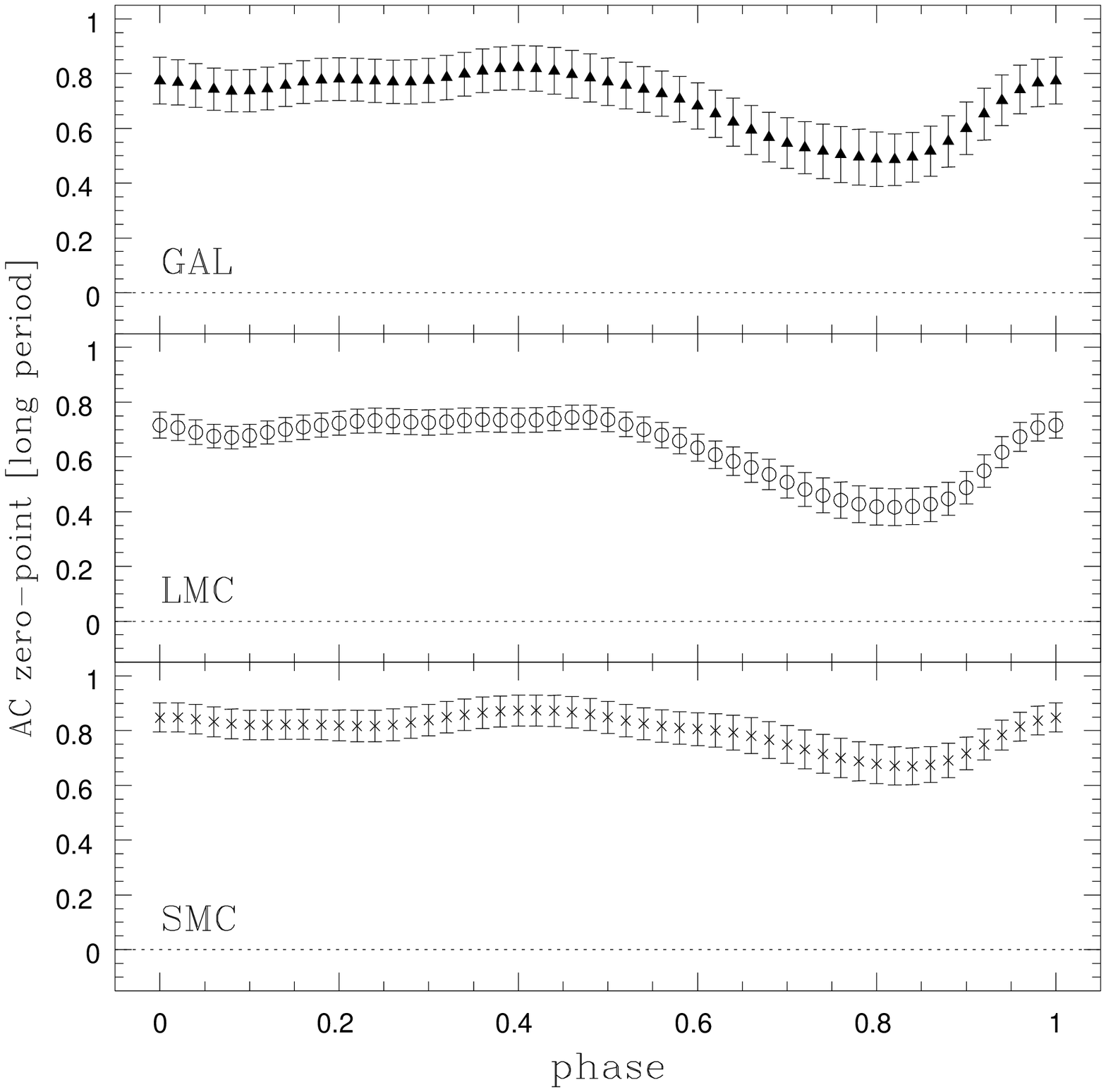}}
       \vspace{0cm}
       \caption{Same as Figure \ref{figpclong}, but for the AC relations.}
       \label{figaclong}
     \end{figure*}

\begin{figure*}
       \vspace{0cm}
       \hbox{\hspace{0.2cm}\epsfxsize=8.5cm \epsfbox{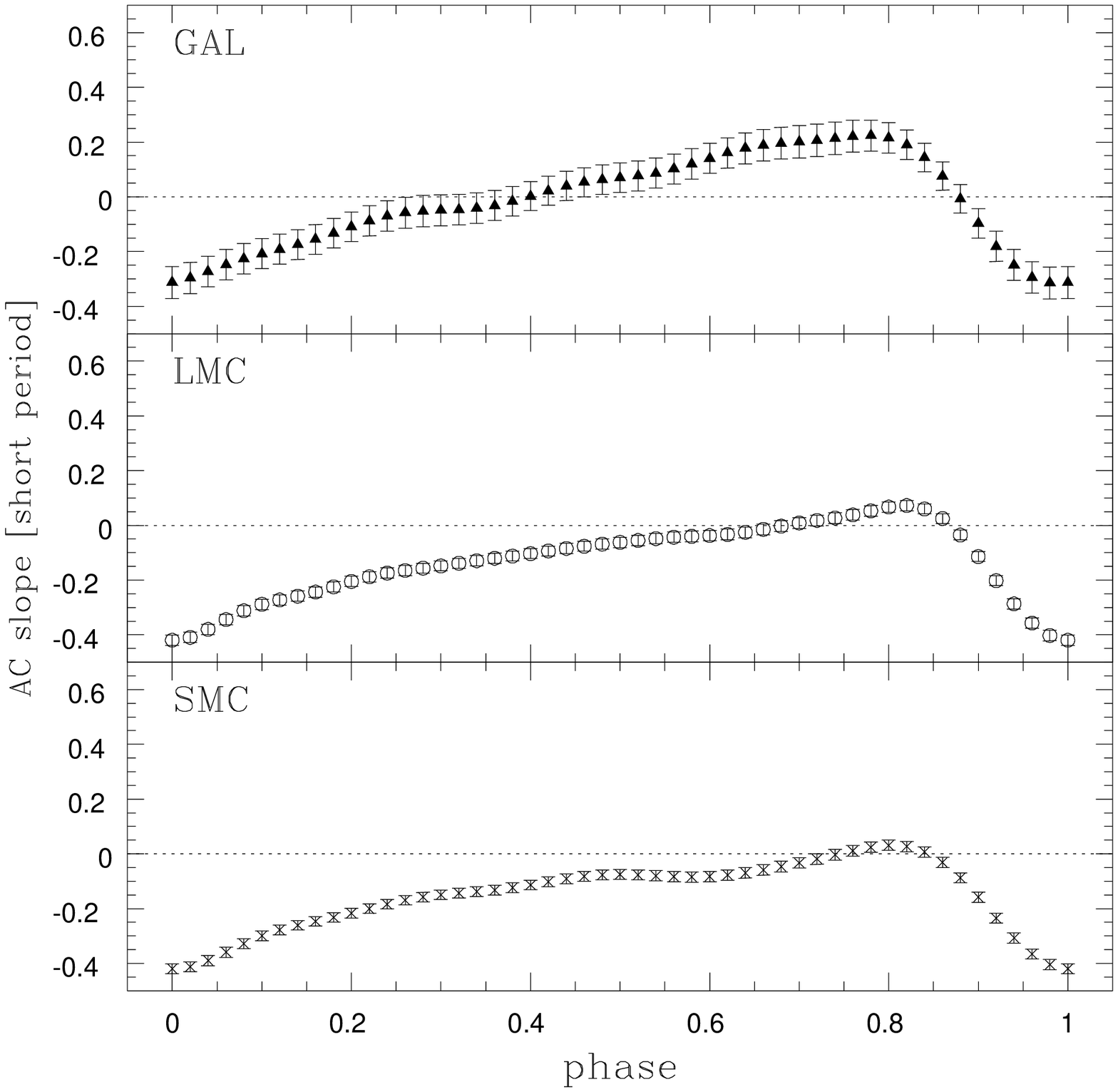}
         \epsfxsize=8.5cm \epsfbox{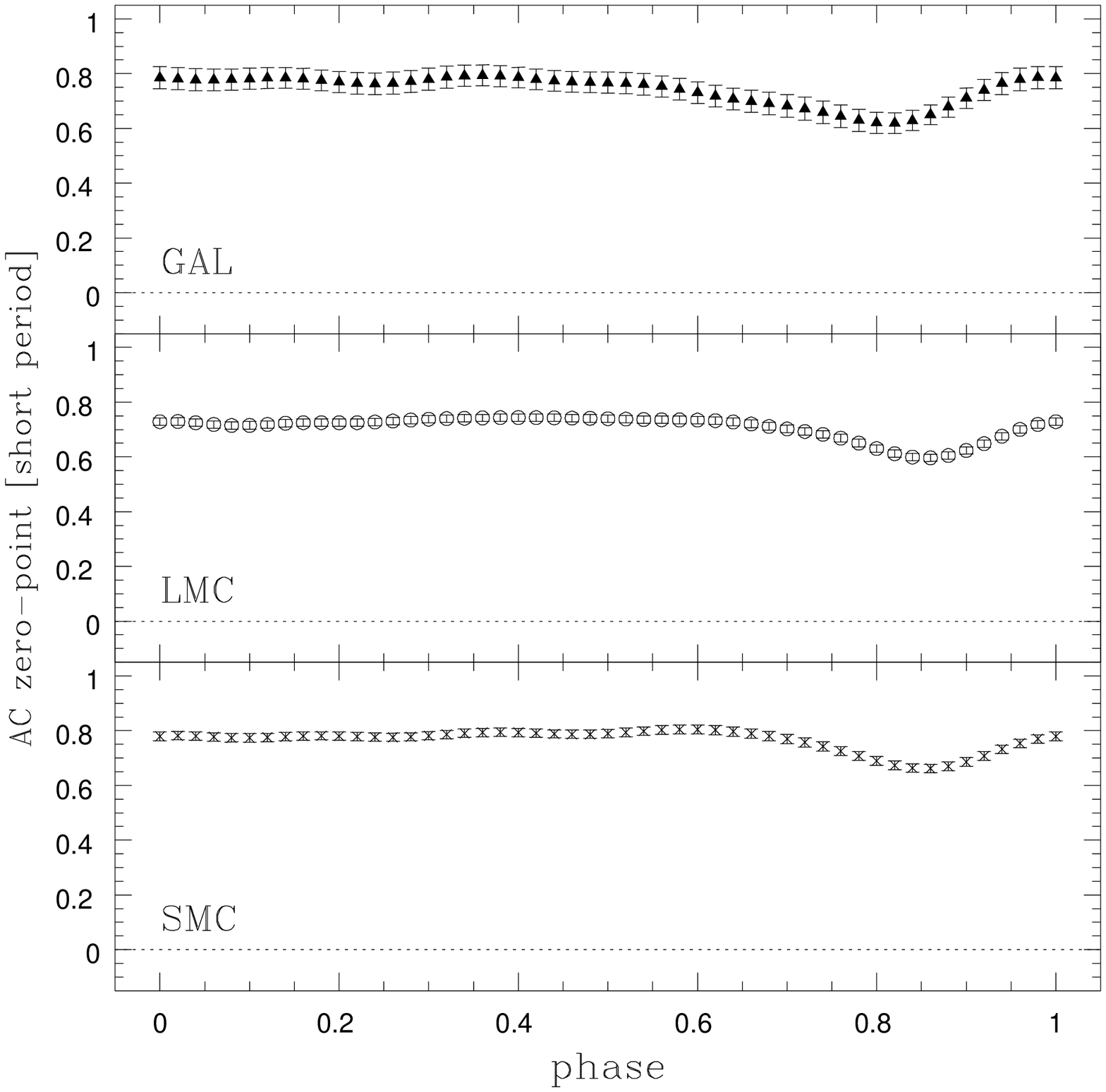}}
       \vspace{0cm}
       \caption{Same as Figure \ref{figpcshort}, but for the AC relations.}
       \label{figacshort}
     \end{figure*}

Figures  \ref{figacall}-\ref{figacshort} present the multi-phase AC relations for the Cepheids in these three galaxies. They support the validity of equation (1) because when the PC slope is close to zero at phases close to maximum light, the AC relation becomes such that higher amplitude stars are driven to cooler temperatures at minimum light. The slope of the AC relation for long period Cepheids reaches a maximum positive value at phases between $0.6-0.8$. The slope stays at this positive value for a phase interval that increases with decreasing metallicity. For phases close to maximum light, the slope is negative for all three galaxies. The slope of the plot depicting the variation of the AC slope with phase for short period Cepheids is similar to the long period plot except that the curves are shifted downward by about 0.2 so that most slopes are negative. 

\section{The Multi-phase LMC PL relations}


\begin{figure*}
       \centering
       \hbox{\hspace{0.25cm}\epsfxsize=7.1cm \epsfbox{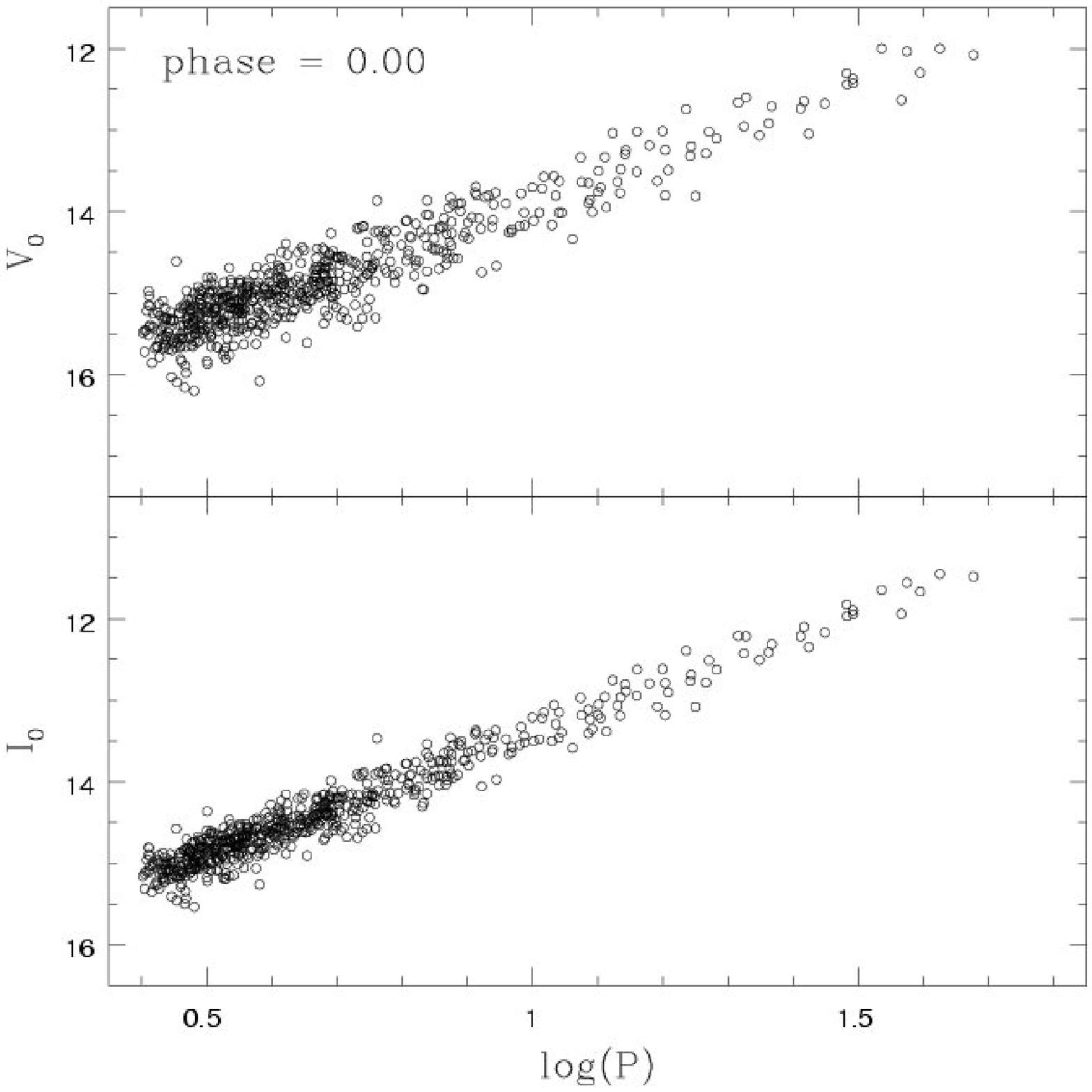}
         \epsfxsize=7.1cm \epsfbox{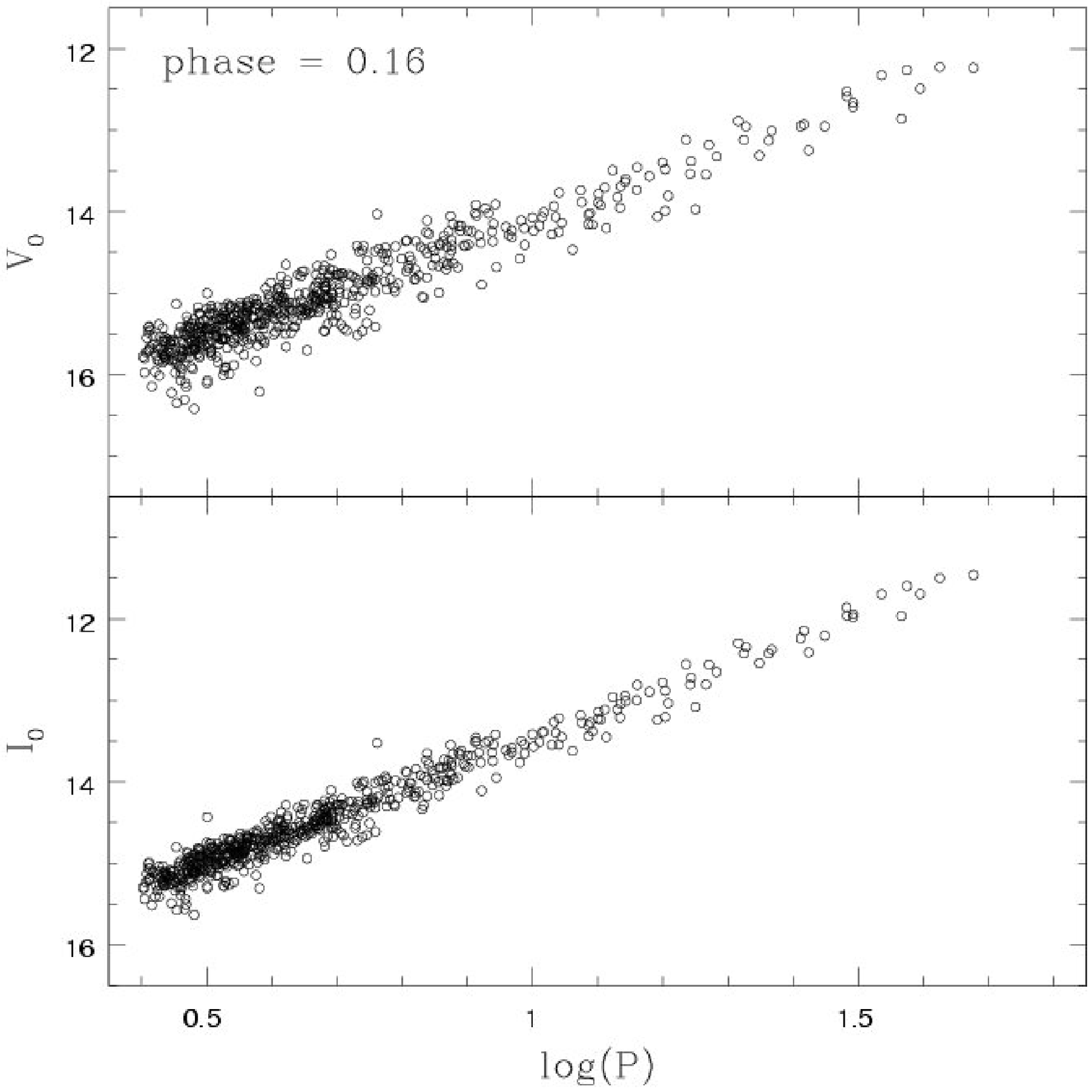}}
       \hbox{\hspace{0.25cm}\epsfxsize=7.1cm \epsfbox{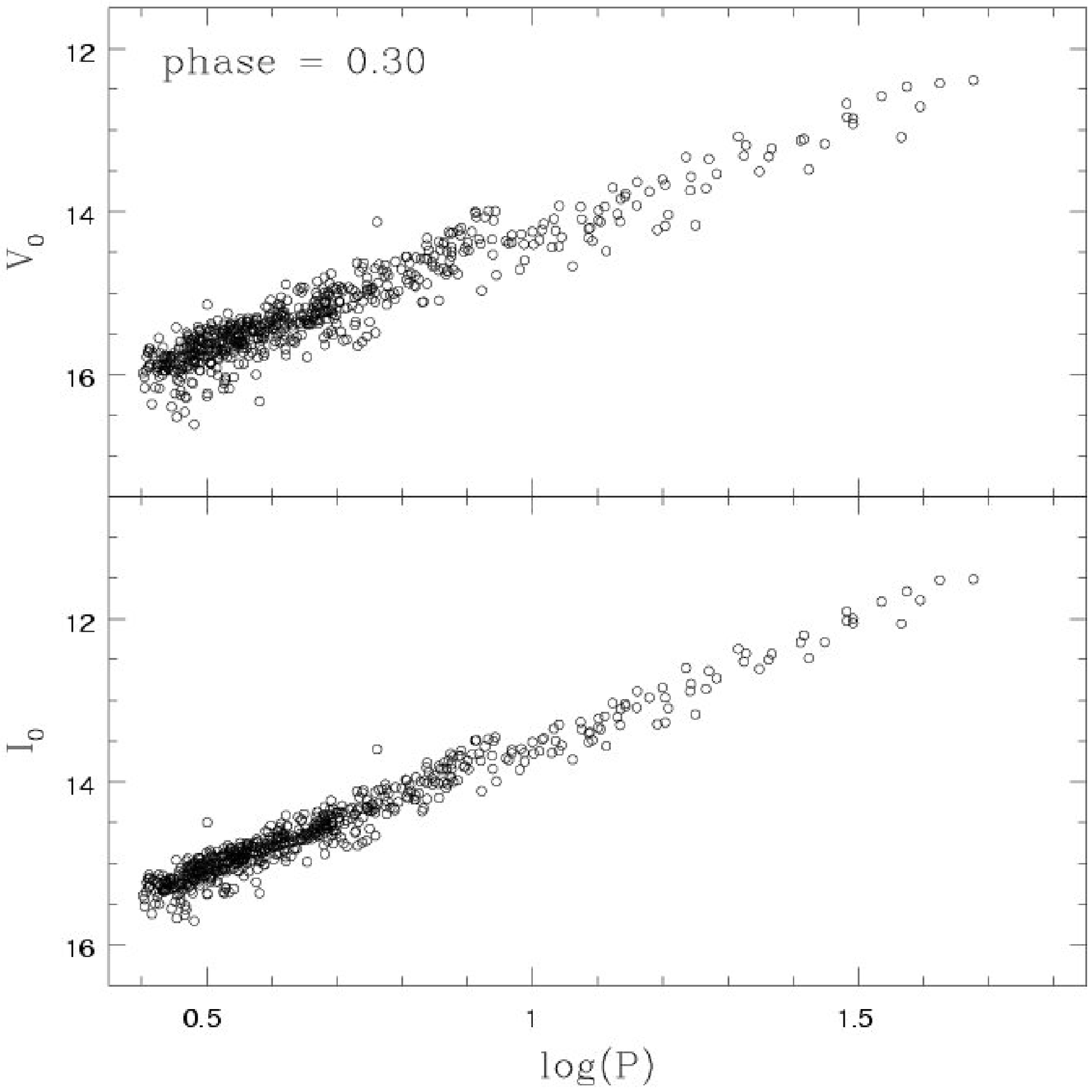}
         \epsfxsize=7.1cm \epsfbox{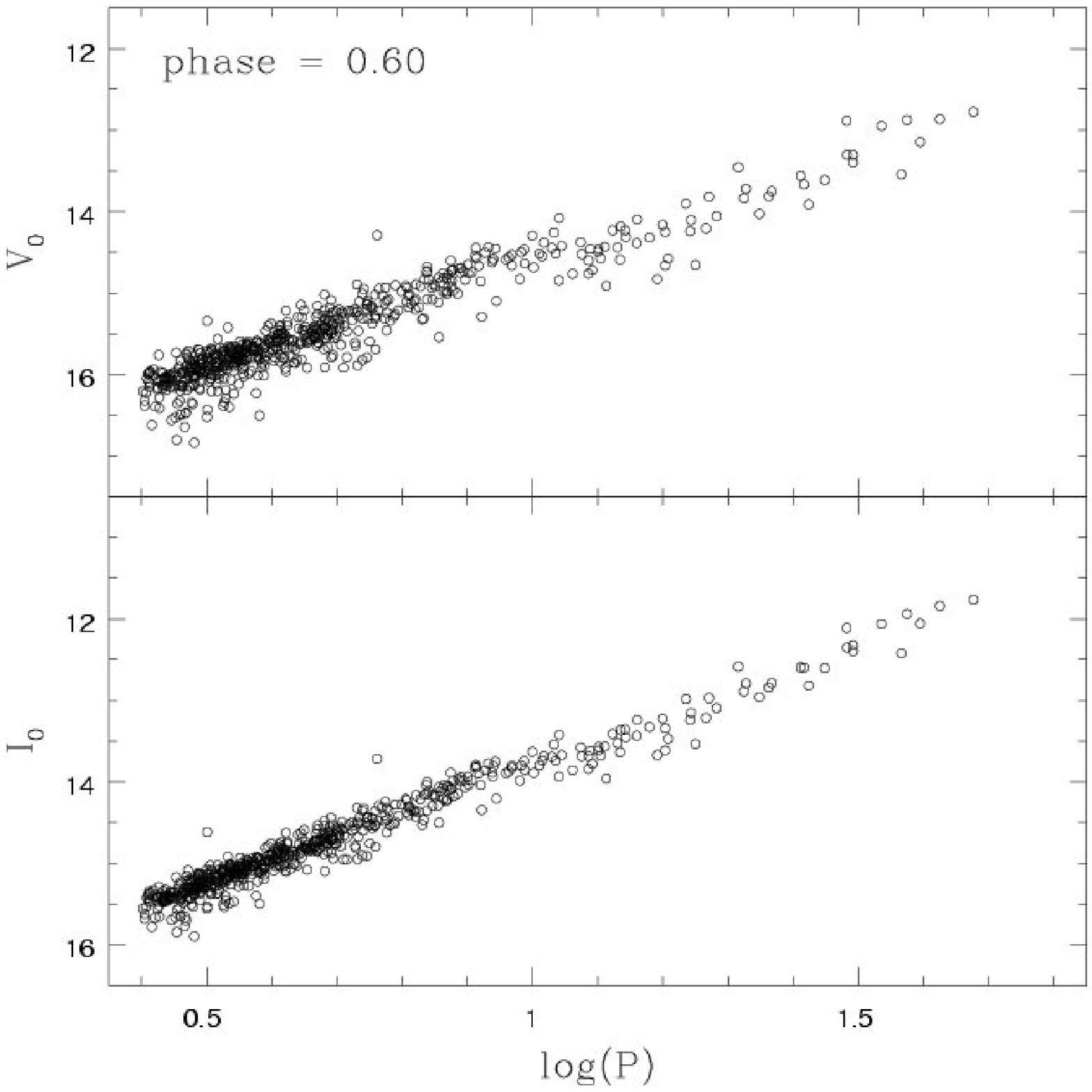}}
       \hbox{\hspace{0.25cm}\epsfxsize=7.1cm \epsfbox{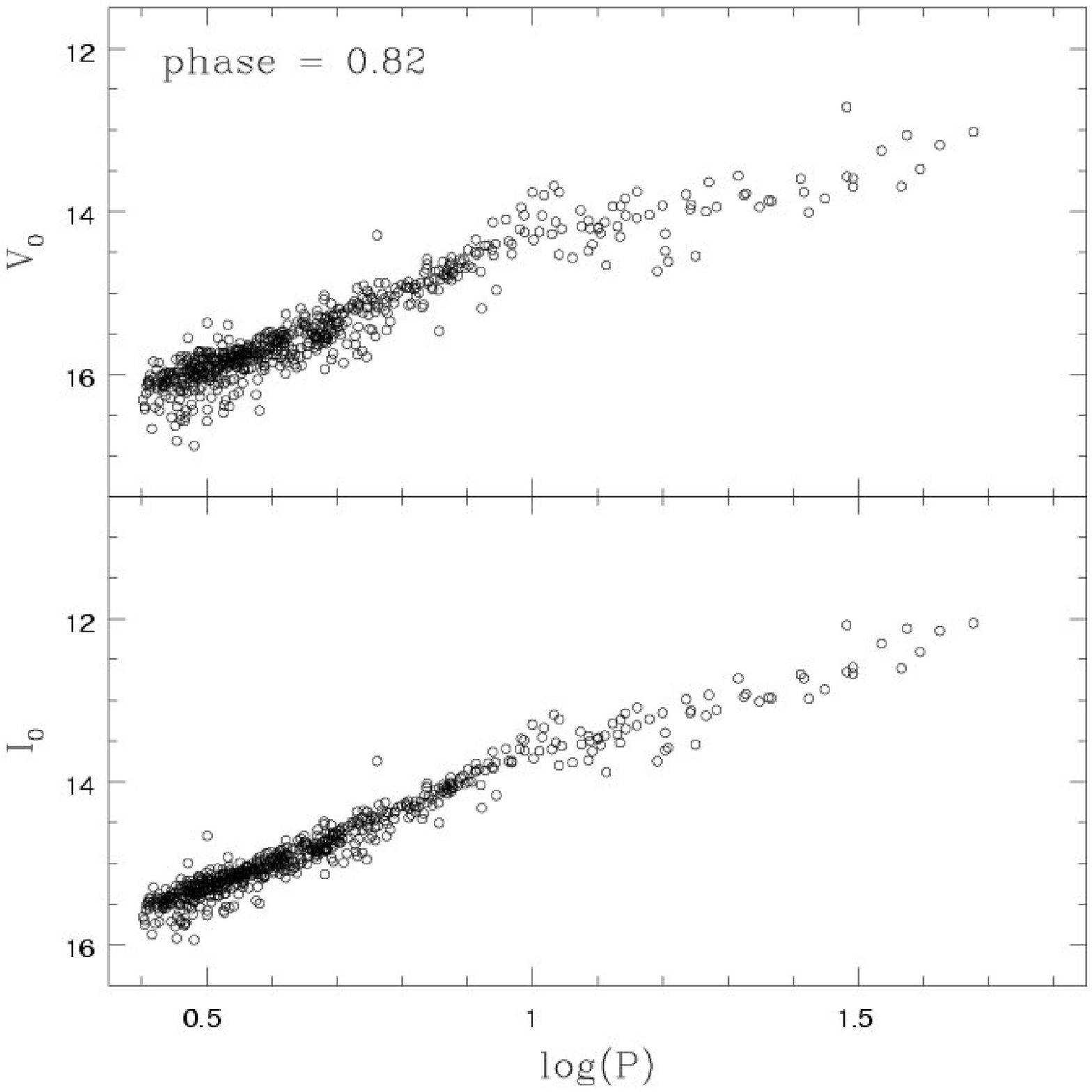}
         \epsfxsize=7.1cm \epsfbox{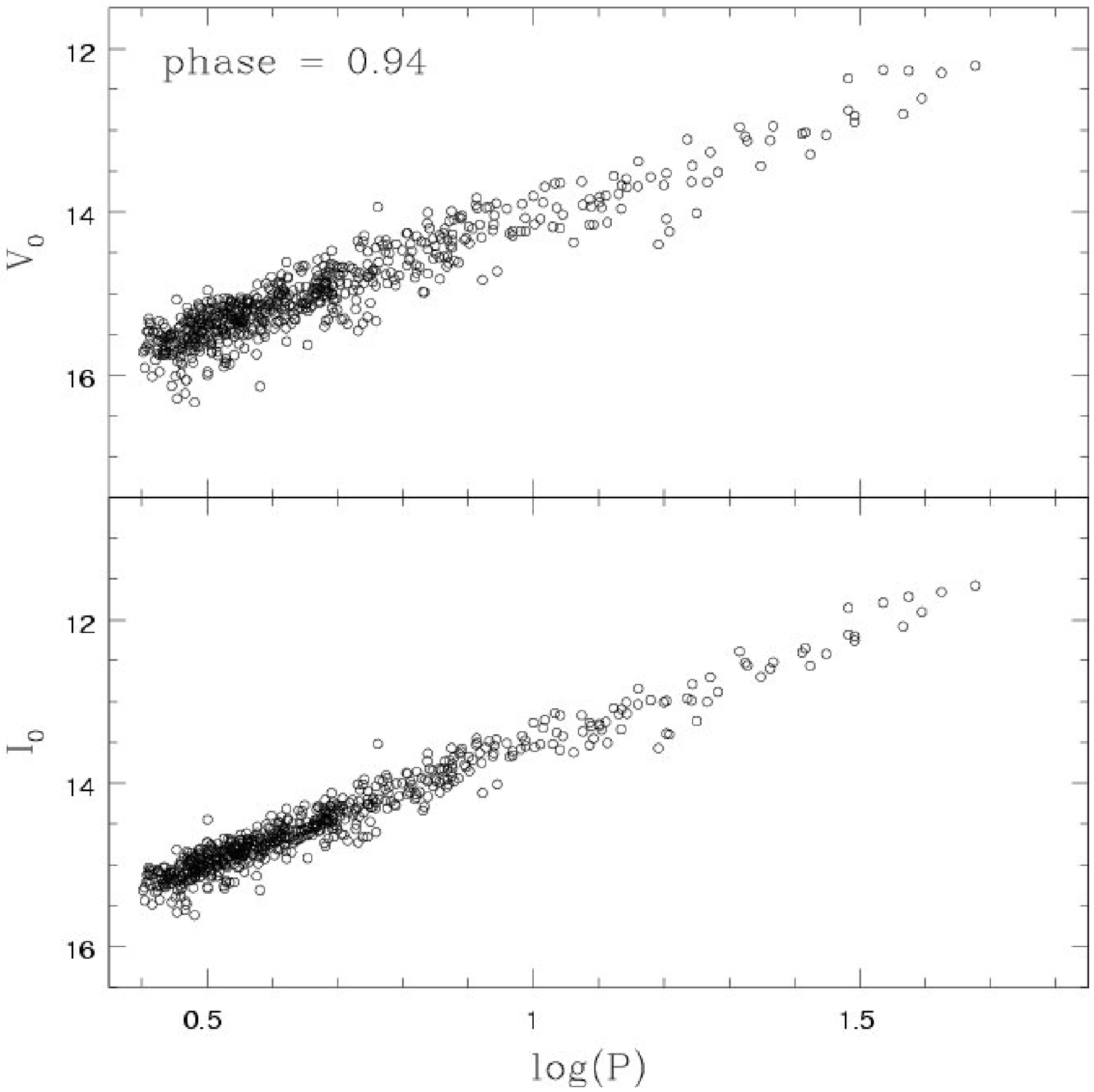}}
       \vspace{0cm}
       \caption{Comparisons of the LMC $V$- and $I$-band PL relations at various phases.}
       \label{plvar}
     \end{figure*}

\begin{figure}
       \hbox{\hspace{0.1cm}\epsfxsize=7.5cm \epsfbox{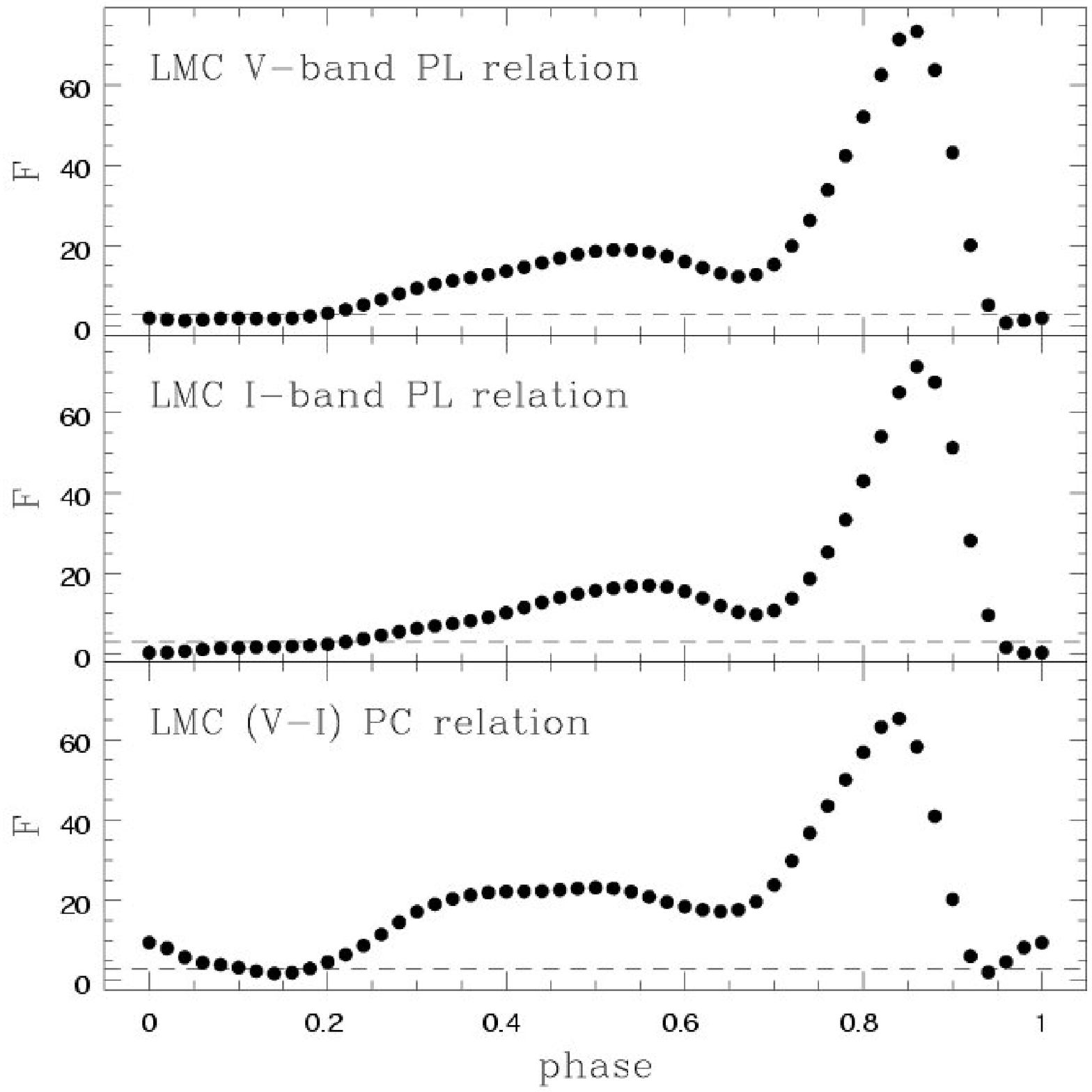}}
       \caption{Comparison of the $F$-values for the LMC PL and PC relation as a function of phase. The dashed lines represent the 95\% confidence level (or $p[F]=0.05$), corresponding to $F\sim3.0$. Hence, the $F$-values above the dashed line indicate that the PL/PC relation is non-linear, while the $F$-values below this line suggest the PL/PC relation is linear.}
      \label{figcom_fpl}
     \end{figure}

     Since the morphology of the PC relations will affect the PL relations, Figure \ref{plvar} portrays these extinction corrected LMC PL relations at the same phases as used in the PC relations in Figure \ref{pcvar}, using the same data and methods mentioned in Section 2. We see clearly that changes in the PC relation are associated with changes in the PL relation: the sharp change in the PL relation in both $V$- and $I$-band at $\Phi=0.82$ is very compelling. It is hard to imagine how this plot could be the result of some combination of period selection/incompleteness/reddening errors \citep[see also][]{nge05}. We also compare the $F$-values of the PL and PC relations as a function of phase from the $F$-test in Figure \ref{figcom_fpl}, similar to Figure \ref{figcom_f}. From this figure it can be seen that the non-linearity of the PC relation traces the non-linearity of the PL relations at most of the phases except near $\Phi\sim0.0$ (corresponding to the maximum light). The discrepancy near $\Phi\sim0.0$ is due to the flat PC relation at maximum light for the long period Cepheids that forces the $V$-band PL slope to be similar to the $I$-band PL slopes\footnote{If the PC(max) relation is not flat for the long period Cepheids, then we expect the $F$-test results near $\Phi\sim0.0$ would not show any non-linearity.}.


\section{Conclusion}

In this paper we have used the excellent phase resolution of OGLE data, combined with existing Galactic data to present multi-phase PC and AC relations for Galactic, LMC and SMC Cepheids. The main conclusions from this study are summarized below.

\begin{enumerate}
\item By examining the slopes of the PC relation as a function of phase, our results confirm that the Galactic PC(mean) relation is linear and the LMC PC(mean) relation is non-linear in the sense that the LMC PC relation is more consistent with two linear relations, one for short and long period Cepheids respectively.

\item Both short and long period Cepheids exhibit interesting new features on these multi-phase plots. The slopes of the PC relation for long period Cepheids reach maximum at a phase around $0.7-0.8$, close to the phase of minimum light. In contrast the slope of the PC relation for short period Cepheids displays a minimum (close to zero) around $\Phi \sim 0.8$. These are the new observational features that have not be reported in the literature, and can be used to further constrain stellar pulsation/evolution models of Cepheids. 

\item Combining these behaviors of the slopes at $\Phi \sim0.8$ for the long and short period Cepheids, the PC relation becomes strongly non-linear at this phase. Further, the Galactic and LMC long period Cepheids display a flat PC relation at phases close to the maximum light (at $\Phi=0.0$). This also cause both of the Galactic and LMC PC relation to be non-linear at maximum light. There are some indications that the extent of this flatness may be metallicity dependent since the SMC has a non-zero PC slope around maximum light. 

\item The reason why previous workers found a non-linear and linear LMC and Galactic PC relation is because short and long period LMC Cepheids have a shallower and steeper slope respectively than their Galactic counterparts in most of the phases. In addition, the LMC PC relations are actually non-linear at most pulsational phases. The Galactic PC relations are linear at most pulsational phases.

\item The multi-phase SMC PC relations show some similarities and differences to the Galactic and LMC counterparts. In particular, the non-linear PC relations around $\Phi \sim0.8$ is significant for all three galaxies, and this suggests the non-linearity of the PC relation around this phase is a universal feature. Perhaps it may has to do with the interaction between the hydrogen ionization front (HIF) and the photosphere at outer envelope of Cepheid variables, or even a shock+HIF+photosphere interaction. However a detailed investigation is beyond the scope of this paper.

\item The shapes of the multi-phase AC relations are similar for both long and short period Cepheids, except the curves for short period Cepheids are shifted downward. The multi-phase AC relations also support the validity of equation (1). 

\item The multi-phase LMC PL relations also trace the corresponding multi-phase PC relations (except at maximum light). Evidence of non-linear PL relations at $\Phi \sim0.8$ is also compelling. It is well known that near resonance, the shape of Cepheid light curves changes dramatically and it could be that near this period, the secondary maximum associated with the Hertzsprung progression may become the primary maximum. Perhaps this may be responsible for the sharp change seen at $\Phi \sim0.8$ in the LMC PC/PL relations. However, even if this is the case, it does not negate the observational fact that there are very compelling changes in the LMC PC/PL relations as a function of phase.

\item Since the mean light PC and PL relation is an average over the relations at every phase, the behavior around $\Phi \sim0.8$ has some impact on the mean light relation which makes the LMC PC/PL relation non-linear but the Galactic and SMC PC/PL relations linear at mean light. Our contention is that this effect is reduced in the Galaxy and SMC because of the combined effect of the changes of the HIF-photosphere interaction and the amplitude of the Cepheids.

\end{enumerate}


\section*{acknowledgments}

SMK acknowledges support from  HST-AR-10673.04-A. We thank an anonymous referee for several useful suggestions.
        


\appendix

\section[]{Fourier parameters for the Cepheids data}

    The fitted Fourier parameters for the fundamental mode Cepheids used in this study are presented here. These Fourier parameters are obtained by fitting equation (2) to the observed data using simulated annealing methods as described in \citet{nge03}. Both of the 154 Galactic Cepheids and the 391 SMC Cepheids are fitted with $M=4$-$6$ Fourier expansion (Paper I), while the 641 LMC Cepheids are mainly fitted with $M=4$-$8$ Fourier expansion and some longer period Cepheids the fits are extended to $M=12$ (Paper III). The results are presented in Table A1 - A6 for the $V$- and $I$-band light curves in the three galaxies. The layout of these tables are nearly identical, and they are available in their complete electronic form at the CDS. Here we only show a portion of Table A1 to indicate its form and content. Note that the Fourier phases ($\phi_k$) in these tables are obtained with equation (2), before applying equation (3) to the light curves. 

        \begin{table*}
        \centering
        \caption{$V$-band Fourier parameters for the Galactic Cepheids$^{\mathrm a}$. }
        \label{tab1}
        \begin{tabular}{ccccccccccc} \hline
        Cepheid Name/ID$^{\mathrm b}$  & $\log P$  & $A_0$ & $A_1$ & $\phi_1$ & $A_2$ & $\phi_2$ & $\cdots$ & $A_{6}$ & $\phi_{6}$ & $E(B-V)$ \\         
        \hline \hline
        SZ AQL	& 1.234	& 8.686 & 0.4811 & 1.2348 & 0.1216 & 0.6823 & $\cdots$ & 0.0236 & 3.1093 & 0.552 \\
        U AQL	& 0.847	& 6.475	& 0.3164 & 2.8197 & 0.1122 & 4.2451 & $\cdots$ & 0.0000	& 0.0000 & 0.371 \\
        \hline
        \end{tabular}
        \begin{list}{}{}
        \item   $^{\mathrm a}$ The entire table is available electronically at the CDS. The definition of the parameters are given in equation (2). The values of $\log P$ and $E(B-V)$ are adopted from the literature. Zero entries mean the corresponding orders are not used in the fit.
        \item   $^{\mathrm b}$ Name for the Galactic Cepheids. For Magellanic Cloud Cepheids, this column gives the OGLE or HV ID number.
        \end{list} 
        \end{table*}

\end{document}